# Ground states of Heisenberg spin clusters from a cluster-based projected Hartree-Fock approach

Shadan Ghassemi Tabrizi[1,a,*] and Carlos A. Jiménez-Hoyos[1,†]

[1]*Department of Chemistry, Wesleyan University, Middletown, CT 06459, USA*

[a]*Present address: Department of Chemistry, University of Potsdam, Karl-Liebknecht-Str. 24-25, D-14476, Potsdam-Golm, Germany*

*shadan_ghassemi@yahoo.com

[†]cjimenezhoyo@wesleyan.edu

**Abstract.** Recent work on approximating ground states of Heisenberg spin clusters by projected Hartree-Fock theory (PHF) is extended to a cluster-based ansatz (cPHF). Whereas PHF variationally optimizes a site-spin product state for the restoration of spin- and point-group symmetry, cPHF groups sites into discrete clusters and uses a cluster-product state as the broken-symmetry reference. Intracluster correlation is thus already included at the mean-field level and intercluster correlation is introduced through symmetry projection. Variants of cPHF differing in the broken and restored symmetries are evaluated for ground states and singlet-triplet gaps of antiferromagnetic spin rings for various cluster sizes, where cPHF in general affords a significant improvement over ordinary PHF, although the division into clusters lowers the cyclical symmetry. On the other hand, certain two- or three-dimensional spin arrangements permit cluster groupings compatible with the full spatial symmetry. We accordingly demonstrate that cPHF yields approximate ground states with correct spin and point-group quantum numbers for honeycomb lattice fragments and symmetric polyhedra.

## 1. Introduction

The calculation of magnetic properties of exchange-coupled spin clusters, e.g., molecules with multiple open-shell transition-metal centers bridged by diamagnetic ligands [1,2], from the Heisenberg model, $\hat{H} = \sum_{i<j} J_{ij} \hat{\mathbf{s}}_i \cdot \hat{\mathbf{s}}_j$, usually relies on approximations, because exact diagonalization (ED) is only feasible for small systems. The ground state and perhaps a few excited states are needed to interpret electron-paramagnetic resonance (EPR) or inelastic neutron scattering (INS) spectra, or to assess other low-temperature properties [3]. The density matrix renormalization group (DMRG [4]) is the most important variational method for ground states of one-dimensional (1D) systems (rings or chains), but is less suitable for 2D coupling topologies. The scarcity of computationally affordable and easily applicable alternatives motivated our recent exploration of projected Hartree-Fock theory (PHF [5]) for ground states of Heisenberg spin clusters [6]. PHF can be used in a black-box manner and has a mean-field (HF) scaling, with a prefactor depending on the size of the symmetry-projection grid. In finite spin systems, PHF restores spin (S) and point-group (PG) symmetry from a general product state. For a collection of $s = \frac{1}{2}$ sites, this broken-symmetry reference state is simply a three-dimensional spin configuration [6]. PHF yields rather accurate ground-state wave functions for symmetric rings with a moderate number of sites $N$ and large local spin $s$ and predicts reasonably accurate singlet-triplet gaps. Limitations become evident for larger rings, where the accuracy decreases [6]. PHF indeed recovers zero correlation energy per site in the thermodynamic limit $N \to \infty$. In other words, the method is not size extensive [7]. To ameliorate this problem and enable a more accurate treatment of larger systems by variational symmetry-projection methods, one could either adopt a multi-component ansatz, where the broken-symmetry reference is a linear combination of non-orthogonal mean-field states [8], or work with a cluster basis that grants more flexibility than ordinary PHF, while still optimizing just a single reference. We pursue the latter option, which we call cPHF. Note however that both strategies could be combined into a multi-component cPHF ansatz, which may be pursued in future work. For other correlated spin-cluster approaches (coupled-cluster and many-body perturbation theory) and for further literature on related methods, see, e.g., Ref. [9].

## 2. Theory and Computations

PHF optimizes a broken-symmetry mean-field state $|\Phi\rangle$ for the application of a symmetry projector $\hat{P}$ [5]. In cPHF, $|\Phi\rangle$ is a product of individual cluster states $|\Phi_i\rangle$,

$$|\Phi\rangle = \prod_{i=1}^{Q} |\Phi_i\rangle, \tag{1}$$

where $Q$ is the total number of clusters. As an example, for a cluster comprising two $s = \frac{1}{2}$ sites, the structure of $|\Phi_i\rangle$ is given in Eq. (2).

$$|\Phi_i\rangle = c_{i,1}|\downarrow\downarrow\rangle + c_{i,2}|\downarrow\uparrow\rangle + c_{i,3}|\uparrow\downarrow\rangle + c_{i,4}|\uparrow\uparrow\rangle \tag{2}$$

The $|\Phi_i\rangle$ are independently optimized to minimize the variational energy, Eq. (3), of the projected state $|\Psi\rangle = \hat{P}|\Phi\rangle$,

$$E = \frac{\langle \Psi | \hat{H} | \Psi \rangle}{\langle \Psi | \Psi \rangle} = \frac{\langle \Phi | \hat{H}\hat{P} | \Phi \rangle}{\langle \Phi | \hat{P} | \Phi \rangle} \quad . \tag{3}$$

The site-permutation invariance [10] of spin Hamiltonians representing systems with spatial symmetry (rings, symmetric polyhedra, etc.) is here referred to as point-group (PG) symmetry. Each level is thus characterized by its total spin $S$ and its PG-label $\Gamma$. To recover a substantial fraction of the correlation energy for all but the smallest systems, it is mandatory to combine S- with PG-projection in PHF [6,11]. In the PG-projector $\hat{P}_\Gamma$ of Eq. (4) (we consider only one-dimensional irreducible representations $\Gamma$), $h$ is the order of the group, $\chi_\Gamma(g)$ is the character of group element $g$, and $\hat{R}_g$ is the respective symmetry operation [12].

$$\hat{P}_\Gamma = \frac{1}{h} \sum_{g=1}^{h} \chi_\Gamma^*(g) \hat{R}_g \tag{4}$$

Multidimensional irreducible representations become relevant for projection onto $S > 0$ sectors. The projector $\hat{P}_m^S$ for spin $S$ and magnetic quantum number $m$ (the $\hat{S}_z$ eigenvalue) is expanded in terms of transfer operators $\hat{P}_{mk}^S$,

$$|\Psi_m^S\rangle = \hat{P}_m^S |\Phi\rangle = \sum_k f_k \hat{P}_{mk}^S |\Phi\rangle \quad , \tag{5}$$

which are conveniently parameterized by Euler angles [13],

$$\hat{P}_{mk}^S = \frac{2S+1}{8\pi^2} \iiint d\alpha d\beta d\gamma \sin(\beta) D_{mk}^{S*}(\alpha,\beta,\gamma) e^{-i\alpha \hat{S}_z} e^{-i\beta \hat{S}_y} e^{-i\gamma \hat{S}_z} \quad . \tag{6}$$

For a given $|\Phi\rangle$, the coefficients $f_k$ [Eq. (5)] correspond to the lowest-energy solution of the generalized eigenvalue problem for the Hamiltonian $\hat{H}$ in the non-orthogonal basis spanned by $\{\hat{P}_{mk}^S |\Phi\rangle\}$, $k = -S, -S+1, ..., +S$ [11]. For combined S- and PG-projection, the projector is a product, $\hat{P} = \hat{P}_m^S \hat{P}_\Gamma$ (spin rotations commute with site permutations). In the trivial case where the projector is the identity, $\hat{P} = \hat{1}$, that is, if no symmetry projection is performed, cPHF is equivalent to cHF, also called cluster mean-field theory [9,14]. If each cluster comprises just one site, cPHF becomes equivalent to PHF, specifically, the "single fermion" variety of PHF presented in Ref. [6]. Finally note that cPHF trivially yields the exact ground state in the chosen symmetry sector $(S,\Gamma)$ if all sites are contained in a single cluster or if there are no couplings between clusters.

In quantum-chemical terminology, $|\Phi\rangle$ is of generalized HF type (GHF [15]) if it completely breaks spin symmetry. An unrestricted HF (UHF) state also breaks total spin symmetry (that is, $|\Phi\rangle$ is not an eigenfunction of $\hat{\mathbf{S}}^2$), but conserves $\hat{S}_z$. In a UHF-type reference, each cluster has a defined $z$-projection $m_i$. Different clusters may have different $m_i$, which add up to the total $\hat{S}_z$ eigenvalue, $m = \sum_i m_i$. Compared to complete spin-symmetry breaking in GHF, the number of variational parameters is reduced in UHF. As an example, a general $m_i = 0$ state of an $s = \frac{1}{2}$ dimer is a superposition of only two basis states, Eq. (7),

$$|\Phi_i\rangle = c_{i,1}|\downarrow\uparrow\rangle + c_{i,2}|\uparrow\downarrow\rangle \;. \qquad (7)$$

PHF variants that restore S- or PG-symmetry from a GHF- or UHF-type reference, are called SGHF, PGSUHF, etc. In cPHF, the cluster size $q$ may be appended, e.g., SGHF(2) denotes a cluster-based SGHF calculation with dimers. For a given grouping, the lowest variational energy is obtained when working with the largest symmetry group (PGSGHF). We do not include complex-conjugation symmetry [5,16] in the cPHF scheme, because this involves a more complicated formalism [6,17,18] and has comparatively small effects for Heisenberg systems [6].

Self-consistent field (SCF [5,6,18,19]) and gradient-based optimization (Refs. [11,20], and references cited therein) are two different strategies for the optimization of $|\Phi\rangle$. In the SCF approach, the local cluster states $|\Phi_i\rangle$ result from successively building and diagonalizing an effective Fock matrix for each cluster. We found that reaching SCF convergence is often challenging in cPHF and therefore prefer gradient-based optimization, where each $|\Phi_i\rangle$ is parameterized in terms of a Thouless rotation from an initial guess $|\Phi_i^0\rangle$. Details are provided in the Supplemental Information (SI). With $q$ sites of spin $s$, the number of real variational parameters that define a general Thouless rotation for a single cluster is $N_{\text{var}} = 2[(2s+1)^q - 1]$, leading to a total of $N_{\text{var}} = 2Q \cdot [(2s+1)^q - 1]$ for $Q$ clusters (not counting the $f_k$ coefficients for $S > 0$, cf. Eq. (5)). Note however that the Thouless parameterization, though convenient, is slightly redundant [6] with respect to S-projection from a GHF-type reference $|\Phi\rangle$, because global spin rotations as well as certain gauge transformations of $|\Phi\rangle$ leave the spin-projected state unchanged [21].

The local cluster basis of dimension $(2s+1)^q$ would have to be truncated for large clusters, e.g., by considering a limited number of lowest levels of the intracluster Hamiltonian. As an example, such a scheme could make a treatment of the $Mn_{70}$ or $Mn_{84}$ single-molecule magnets (with $N = 70$ or $N = 84$ $s = 2$ sites) feasible in terms of a $q = 7$ division [22] but is beyond the scope of this work.

As recommended previously [19], we discretized transfer operators, Eq. (6), with a combined Lebedev-Laikov [23] and Trapezoid integration grid. For S-projection from a UHF-type reference, the evaluation of integrals over Euler angles $\alpha$ and $\gamma$ is trivial [24] and integration over $\beta$ employs a Gauss-Legendre grid. A computational parallelization of the summation over the grid is trivial [19]. The quality of S-projection can be checked by computing $\langle \Psi | \hat{\mathbf{S}}^2 | \Psi \rangle$ from a sum of spin-pair correlation functions (SPCFs), Eq. (8),

$$\langle \hat{\mathbf{S}}^2 \rangle = Ns(s+1) + 2\sum_{i<j} \langle \hat{\mathbf{s}}_i \cdot \hat{\mathbf{s}}_j \rangle . \tag{8}$$

We ensured that $\langle \hat{\mathbf{S}}^2 \rangle$ deviates by $<10^{-6}$ from the ideal value of $S(S+1)$. Details on the calculation of SPCFs are provided in SI.

Figure 1 illustrates that cluster formations are in general not fully compatible with spatial symmetry, meaning that working with the cyclic $C_N$ group (or the dihedral group $D_N$ that additionally includes vertical $C_2$ axes) of spin rings with $N$ sites would involve complicated transformations between different cluster bases. A division in terms of $q$ neighbors thus reduces the cyclical symmetry of rings according to $C_N \to C_{N/q}$ or $D_N \to D_{N/q}$. For example, for $q = 2$, sectors $k$ and $(k+\frac{N}{2}) \mod N$ of group $C_N$ [the crystal momentum $k$ indicates the eigenvalue $\exp(-i2\pi k/N)$ of the cyclic permutation $\hat{C}_N$] belong to the same sector of group $C_{N/2}$. Thus, a $k = 0$ state in $C_{N/2}$ is generally a mixture of $k = 0$ (Mulliken label $\Gamma = A$) and $k = \frac{N}{2}$ ($\Gamma = B$) in $C_N$. This however does not imply that cPHF wave functions will significantly break symmetry with respect to the full point group (see Results section).

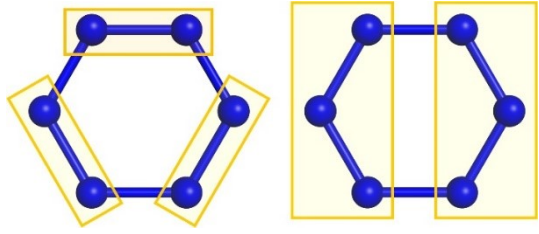

Figure 1: Two cluster groupings in an $N=6$ ring. Dihedral symmetry is broken by dimerization $(D_6 \rightarrow D_3)$ or trimerization $(D_6 \rightarrow D_2)$. The respective reduced symmetry groups can be employed in cPHF.

From the perspective of molecular magnetism, symmetry reduction through cluster formation may not be a major concern, because the cyclic-symmetry order of existing ring-like molecules is often lower than the number of open-shell ions. For example, the mere six-fold rotational symmetry of an $Fe_{18}$ ring (studied by INS in Ref. [25]) suggests a treatment in terms of six equivalent clusters, each hosting three neighboring $s = \frac{5}{2}$ sites, which represent three chemically inequivalent $Fe^{III}$ ions. Furthermore, $Mn_{70}$ and $Mn_{84}$ tori, the largest single-molecule magnets known to date, have repeat units of 14 $Mn^{III}$ ions ($s = 2$), but the pattern of isotropic couplings in the Heisenberg model suggests a partitioning into two inequivalent types of clusters with seven sites each [22].

On the other hand, it is straightforward to use all spatial symmetries in cPHF that keep clusters intact, meaning that a given operation transforms all sites of one cluster into sites of a specific second cluster. Operations associated with internal site permutations do not pose a problem. For example, vertical $\hat{C}_2$ rotations in the dihedral group of rings involve such internal permutations, see SI for details. The Results section additionally presents selected systems (honeycomb lattice fragments and symmetric polyhedra) permitting cluster groupings that are fully symmetry-compatible if operations associated with internal permutations are considered.

Unless noted otherwise, all clusters (total number $Q = N/q$) are equivalent by symmetry and contain the same number of sites $q$ with a uniform local spin $s$, although these are not requirements for the application of cPHF. It is usually reasonable to construct clusters from neighboring (interacting) sites to recover a substantial amount of correlation energy already at the mean-field level. In the absence of intracluster interactions, cHF recovers the ordinary HF (classical spin, $q = 1$) solution. With respect to variational symmetry projection in cPHF, the cluster ansatz offers more flexibility than $q = 1$ (ordinary PHF), even when there are no intracluster interactions (see Results section).

The relative correlation energy, $0 \leq p \leq 1$, defined in Eq. (9),

$$p = \frac{E_{cPHF} - E_{HF}}{E_0 - E_{HF}},\qquad(9)$$

is an accuracy measure of cPHF for ground states, where $E_0$ is the exact result (ED) and $E_{HF}$ refers to the classical solution, e.g., $E_{HF} = -NJs^2$ for the Néel configuration in a ring. All benchmark systems treated here have a nondegenerate $S = 0$ ground state. Energies are reported in units of the uniform nearest-neighbor coupling constant $J = 1$ (antiferromagnetic coupling).

In cases where the system size prohibits a comparison with exact energies, we compare cPHF against a cluster-variant of second-order perturbation theory (cPT2). The (non-variational) cPT2 corrected energy is given by Eq. (10),

$$E_{cPT2} = E_{cHF} - \frac{\sum_{ab}|\langle\Phi_{cHF}|\hat{H}|\Phi_{ab}\rangle|^2}{E_{cHF} - E_{ab}},\qquad(10)$$

where $E_{cHF} = \langle\Phi_{cHF}|\hat{H}|\Phi_{cHF}\rangle$, $E_{ab} = \langle\Phi_{ab}|\hat{H}|\Phi_{ab}\rangle$, and $|\Phi_{ab}\rangle$ is obtained from $|\Phi_{cHF}\rangle$ through excitations $a$ and $b$ in two neighboring (directly interacting) clusters. In other words, $|\Phi_{ab}\rangle$ is a doubly excited cluster mean-field state. The denominator in the second term of Eq. (10) is of Epstein-Nesbet type [26]. For cluster size $q = 1$ and $s = \frac{1}{2}$ such PT2 corrections were considered for the truncated icosahedron in Ref. [27]. Alternatively, one could apply Rayleigh-Schrödinger perturbation theory in a cluster basis, using differences of Fock-like orbital energies in the denominator, as done in Refs. [9] (for the $s = \frac{1}{2}$ square lattice) and Ref. [28] (for the single-band Hubbard model).

## 3. Results and Discussion

For antiferromagnetic $s = \frac{1}{2}$ spin rings, honeycomb lattice fragments, and four tetrahedral or icosahedral polyhedra, we compare ground-state energies and SPCFs from different variants of cPHF against exact results (where available). In addition, we briefly consider singlet-triplet gaps in spin rings and larger local spin (up to $s = 2$) in polyhedra and explore how the quality of predictions depends on the cluster size or shape. Rather than deriving specific new insights on any of these systems, our aim is to investigate the potential of cPHF as a variational black-box method with a cluster mean-field scaling.

**Symmetric rings.** As in our previous work [6], we choose rings with $N = 6, 12, 18, 24, 30$ sites as benchmark systems to compare against ED, but restrict attention to the least classical case $s = \frac{1}{2}$, which had proven to be most problematic for ordinary PHF ($q = 1$) [6].

We note in passing that, in contrast to dimers formed from neighboring sites (see Theory section), clusters composed of diametrically opposite sites are indeed compatible with the full $D_N$ symmetry. Figure 2 illustrates that the cyclic $\hat{C}_N$ operation leaves all clusters intact and exchanges sites in one dimer. For $N = 12$, $C_N$SGHF(2) with such a partitioning yields the exact $S = 0$ ground state (exact within numerical double precision) in each $k$ sector ($k = 0, 1, \ldots, N-1$), whereas $C_N$SGHF(1) turned out to be exact in sectors $k = 1, 3, 5, 7, 9, 11$ only [6]. This shows that the cluster ansatz is somewhat more flexible with respect to variational symmetry projection, even though HF(2) is equivalent to HF(1), due to the lack of intracluster interactions. For larger rings, the described pairing of the most distant sites does not provide a sizable improvement over PHF(1), and shall not be discussed further.

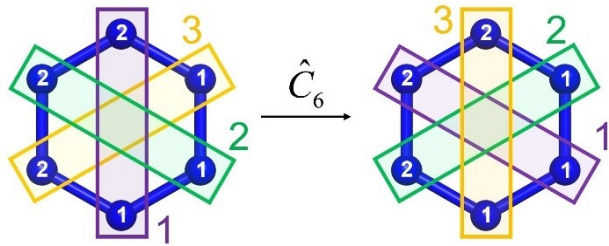

Figure 2: Dimers of diametrically opposite sites conserve the symmetry of rings (group $D_6$ in this example). A cyclic permutation interchanges sites in the last cluster (cluster 3, yellow box).

We construct clusters comprising $q = 2, 3, 6$ neighbors (Figure 3), corresponding to the common divisors of $N$, thus reducing the point group accessible to cPHF, that is, $D_N \to D_Q$, where $Q = N / q$. Rings with even $N$ lack frustration and adopt UHF-type solutions in cHF. In contrast, for odd $N$ (not studied here) frustration gives rise to genuine GHF solutions breaking $\hat{S}_z$ symmetry.

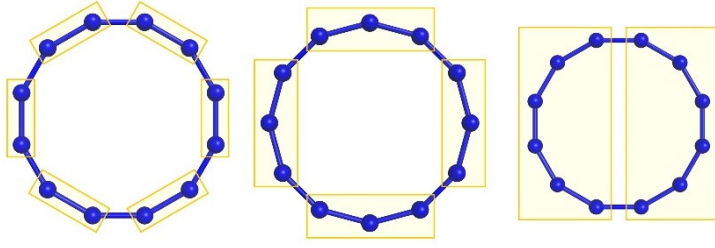

Figure 3: Clusters of size $q = 2, 3, 6$ in an $N = 12$ ring.

It is simple to show that HF(2) produces a product of singlets, $s_i = 0$. We observed that local singlets are also formed in HF(6). Consequently, $E_{HF}$ is a sum of the ground-state energies of the individual clusters: $E_{HF(2)} = -\tfrac{3}{4} Q$, and $E_{HF(6)} = -2.4936 Q$ (for $Q > 1$). On the other hand, for $q = 3$, all clusters assume $z$-projections $|m_i| = \tfrac{1}{2}$ with alternating signs, leading to a Néel-like spin-density pattern and non-vanishing interactions between clusters, $E_{HF(3)} = -1.1284 Q$. We note in passing that for $s = 1$ and even $q$, all clusters have the same $m_i = 0$ state in HF, but $s_i \neq 0$ and the intercluster interaction is non-zero; for odd $q$, states alternate between $m_i = +1$ and $m_i = -1$.

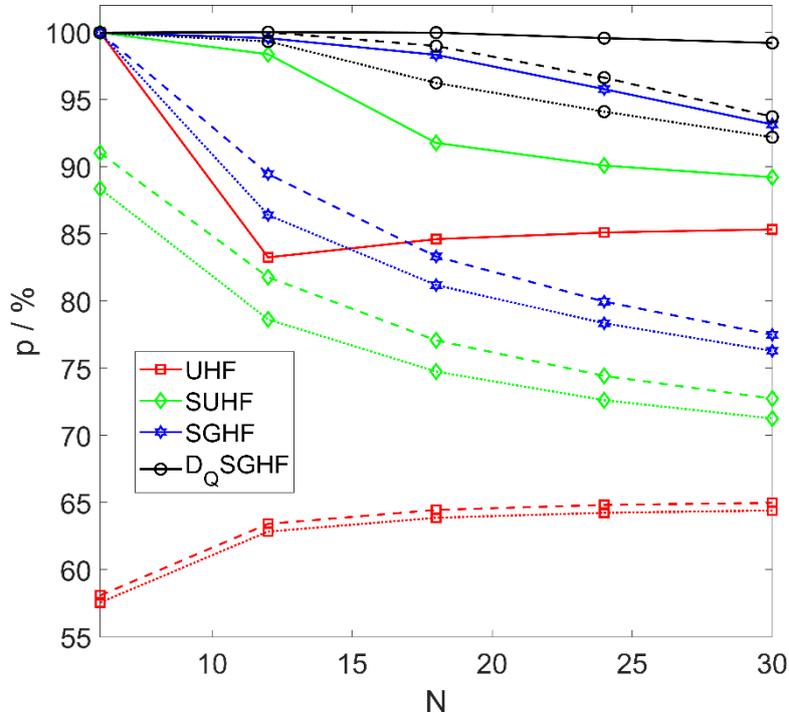

Figure 4: Relative correlation energies $p$ (Eq. (9)) from cluster variants of UHF, SUHF, SGHF and $D_Q$SGHF ($Q = N/q$) in antiferromagnetic spin rings with $N = 6, 12, 18, 24, 30$ sites. Data points are connected by dotted ($q = 2$), dashed ($q = 3$) or solid lines ($q = 6$).

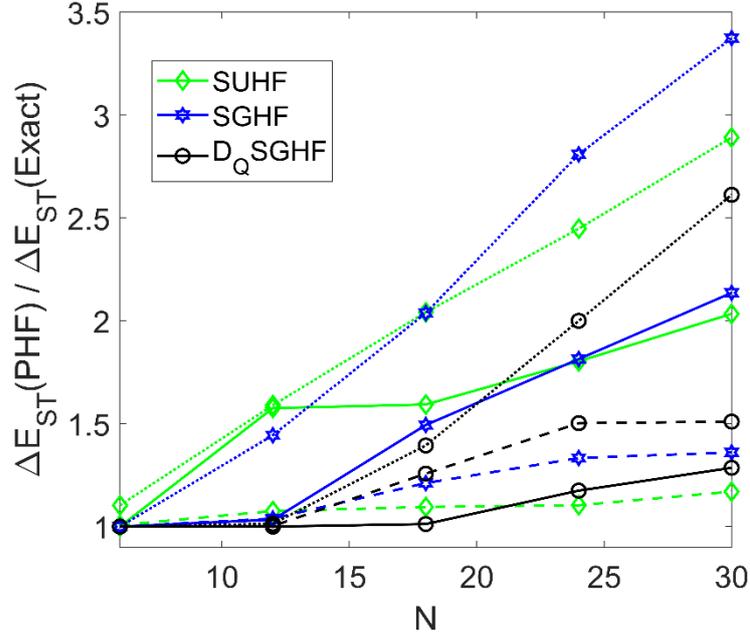

Figure 5: Relative singlet-triplet gaps from cluster variants of SUHF, SGHF and $D_Q$SGHF ($Q = N/q$) in antiferromagnetic spin rings with $N = 6, 12, 18, 24, 30$ sites. Data points are connected by dotted (cluster size $q = 2$), dashed ($q = 3$) or solid lines ($q = 6$).

Relative correlation energies $p$ for cluster variants of UHF, SUHF, SGHF and $D_Q$SGHF are plotted in Figure 4. Relative singlet-triplet gaps from SUHF, SGHF, $D_Q$SGHF and ED are plotted in Figure 5. Numerical energy values are given for reference in Section 3 of SI. The $D_Q$ labels $\Gamma$ for the ground states in sectors $S = 0$ and $S = 1$ depend on $q$, $S$ and $N$, as detailed in

Table 1 (this information can be derived straightforwardly from the respective labels in $D_N$, which were given in Table III of Ref. [6]).

Table 1: Mulliken labels[a] of ground states in sectors $S = 0$ and $S = 1$ of even $N$ antiferromagnetic $s = \frac{1}{2}$ rings in the reduced dihedral group $D_Q$ ($Q = N/q$).

| $N = 4n+2$ | | | | $N = 4n$ | | | |
|---|---|---|---|---|---|---|---|
| $S = 0$ | | $S = 1$ | | $S = 0$ | | $S = 1$ | |
| $q$ even | $q$ odd | $q$ even | $q$ odd | $q$ even | $q$ odd | $q$ even | $q$ odd |
| $A_2$ | $B_1$ | $A_1$ | $A_1$ | $A_1$ | $A_1$ | $A_2$ | $B_1$ |

[a]Label A(B) and subscript 1(2) respectively denote symmetry(antisymmetry) under the cyclic permutation $\hat{C}_Q$ or the vertical $\hat{C}_2$ operation that exchanges all sites pairwise.

For $q=2$ or $q=6$, all clusters are $m_i=0$ in the SUHF references for $S=0$ or $S=1$ projection. The fact that $S=0$ energies are lower than UHF shows that SUHF does not yield a singlet-product. For a given state, system and $q$, the ordering $E_{\text{UHF}} \geq E_{\text{SUHF}} \geq E_{\text{SGHF}} \geq E_{\text{PGSGHF}}$ reflects the breaking and restoration of additional symmetries. (On the other hand, in the absence of a group/subgroup relation between cPHF variants, such as in the set SGHF, PGGHF or PGSUHF, it is not possible to determine the ordering *a priori*.) The equality holds only when the lower-level method is exact. This obtains for $q=6$, $N=6$, where UHF (and any PHF variant) is trivially exact, as well as in a few other cases, e.g., SGHF(2) yields the numerically exact $S=0$ and $S=1$ ground states for $N=6$.

For $q=2$ or $q=3$, an SUHF or SGHF wave function can be trivially expressed in the $q=6$ basis. Thus, $q=6$ will yield the best variational energy. On the other hand, if $D_Q$-projection is included, or if the larger cluster size is not a multiple of the smaller size, we cannot rigorously predict if a larger $q$ is advantageous. However, Figure 4 shows that this is indeed true in all studied cases. Excluding the exact $q=6$, $N=6$ point, the relative correlation $p$ captured by UHF increases with $N$, because rings with even $N$ approach the Bethe-Hulthén limit $\lim_{N\to\infty} E/N = \frac{1}{4} - \ln 2$ from below [29]. With $E_{\text{HF}}/N = -\frac{1}{4}$, the limits for cHF in Eqs. (11)–(13),

$$\lim_{N\to\infty} p_{\text{UHF}(2)} = (\tfrac{1}{4} - \tfrac{3/4}{2})/(\tfrac{1}{2} - \ln 2) \approx 64.7\% , \qquad (11)$$

$$\lim_{N\to\infty} p_{\text{UHF}(3)} \approx (\tfrac{1}{4} - \tfrac{1.1284}{3})/(\tfrac{1}{2} - \ln 2) \approx 65.3\% , \qquad (12)$$

$$\lim_{N\to\infty} p_{\text{UHF}(6)} \approx (\tfrac{1}{4} - \tfrac{2.4936}{6})/(\tfrac{1}{2} - \ln 2) \approx 85.7\% , \qquad (13)$$

also hold for cPHF, due to the lack of size extensivity (see Introduction). This limitation of cPHF is clearly apparent from the fact that $p$ decreases with increasing $N$ (Figure 4), but the improvement over ordinary PHF (where $\lim_{N\to\infty} p = 0$ for $q=1$) is still significant. For example, for $N=30$, $D_{30}$SGHF(1) ($N_{\text{var}}=60$) yields $E=-11.814$ [6] versus $E=-13.276$ with $D_5$SGHF(6) ($N_{\text{var}}=630$). Finally, $E=-13.320$ from $D_3$SGHF(10) ($N_{\text{var}}=6138$; not included in Figure 4) is very close to $E_0=-13.322$ [30].

Triplet energies improve along the same hierarchy as singlet energies, but this does not guarantee an improvement in the gap $\Delta E_{ST}$. Somewhat unfortunately, $\Delta E_{ST}$ is severely overestimated, except for small $N$ and large $q$. The cluster approach can still afford better results than ordinary PHF. For example, for $N = 30$, $D_{30}$SGHF(1) yields $\Delta E_{ST} = 0.212$ [6], compared to $\Delta E_{ST} = 0.189$ from $D_5$SGHF(6) and an exact value of $\Delta E_{ST} = 0.147$ [30].

**Honeycomb lattice fragments.** Systems with the connectivity of polycyclic aromatic hydrocarbons – triphenylene ($N$ = 18), coronene ($N$ = 24), hexabenzocoronene ($N$ = 42), hexa-*cata*-hexabenzocoronene ($N$ = 48), and kekulene ($N$ = 48) – allow to briefly explore options for cluster formation (the chemical nomenclature for these systems is not meant to imply that the Heisenberg model describes properties of the respective organic molecules). The lattices are bipartite [31] and thus have $S$ = 0 ground states. For $s = \frac{1}{2}$ and up to eight different groupings (Figure 6–Figure 10; in some cases, the clusters are obviously not all equivalent), Table 2 compares energies from cUHF, cPHF and cPT2 to exact results (available only for $N$ = 18, 24).

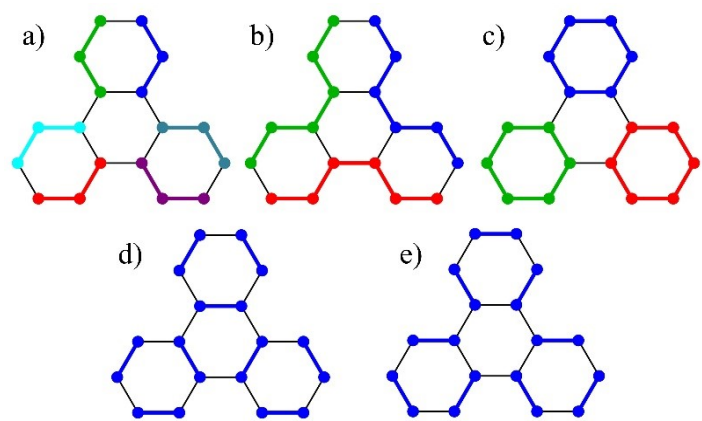

Figure 6: Cluster groupings in triphenylene ($N$ = 18) conserving the full $D_3$ symmetry. In (d) and (e), all dimer clusters ($q$ = 2) are shown in the same color, for simplicity.

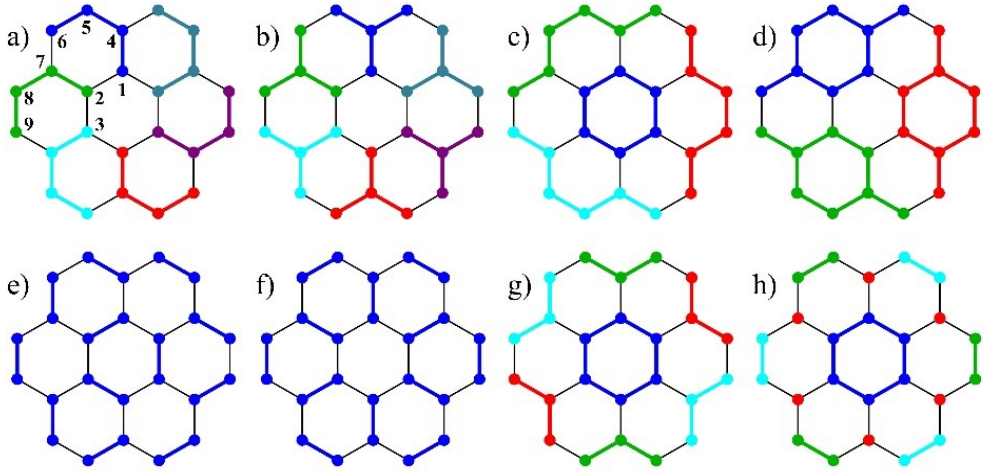

Figure 7: Cluster groupings in triphenylene ($N = 24$) with symmetries $C_6$ (a), $D_6$ (b, f, g, h), and $D_3$ (c, d, e). In (e) and (f), all dimers ($q = 2$) are shown in the same color, for simplicity. In (g), three clusters comprise two separate trimers each. In (h), two clusters comprise three separate dimers, and six disconnected sites (red) constitute another cluster.

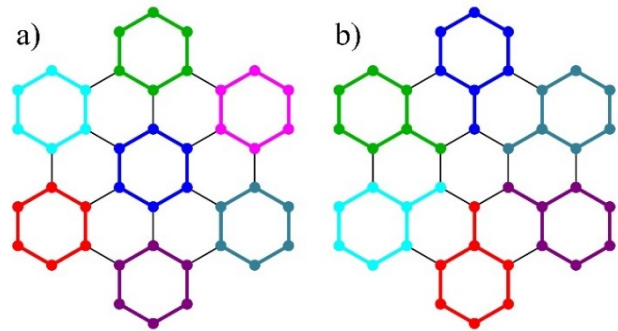

Figure 8: Cluster groupings in hexabenzocoronene ($N = 42$) conserving the full $D_6$ symmetry.

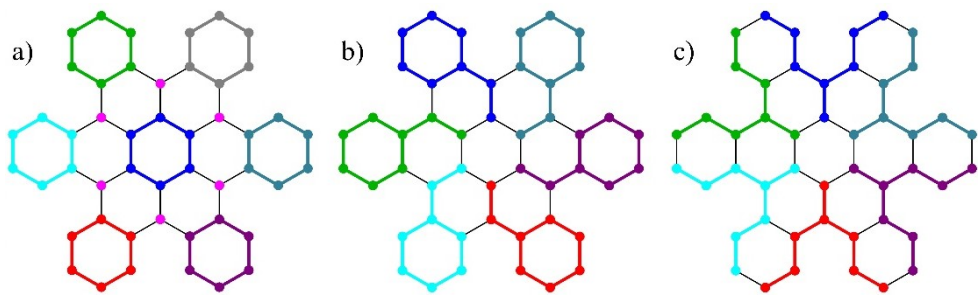

Figure 9: Cluster groupings in hexa-*cata*-hexabenzocoronene ($N = 48$) have symmetries $D_6$ (a, c) or $C_6$ (b). In (a), six disconnected sites (pink) constitute one cluster.

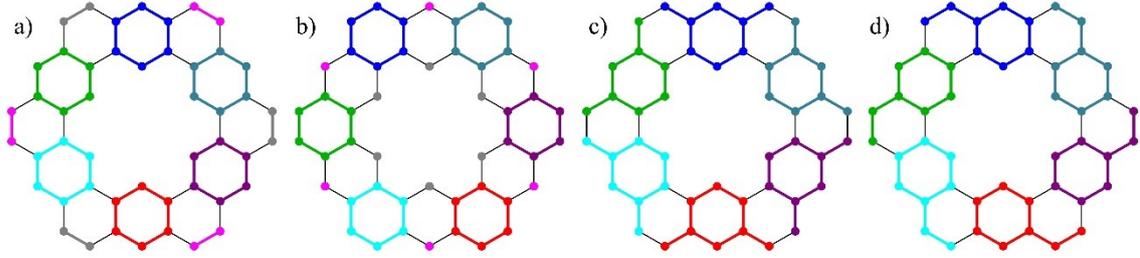

Figure 10: Cluster groupings in kekulene ($N = 48$) with symmetries $D_3$ (a), $D_6$ (b, c), and $C_6$ (d). In (a) and (b), two clusters (grey and pink) each comprise three separate dimers (a) or six separate sites (b).

Table 2: Ground-state energy estimates for honeycomb-lattice fragments from cluster variants of UHF, SGHF, PGSGHF and PT2. The lowest energy for each grouping/method is given in bold type.

| System | Grouping ($q$, bonds)[a] | UHF | SGHF | PGSGHF ($\Gamma$) | PT2 | Exact |
|---|---|---|---|---|---|---|
| Triphenylene | a (3, 12) | -7.2753 | -8.2062 | -8.6556 ($A_2$) | -8.6445 | -8.7697 |
| | b (6, 15) | -7.5804 | -8.4865 | -8.7342 ($A_2$) | -7.8205 | |
| | c (6, 18) | **-8.4083** | **-8.7556** | **-8.7696** ($A_2$) | **-8.6640** | |
| | d (2, 9) | -7.0229 | -8.0255 | -8.5364 ($A_2$) | -8.2068 | |
| | e (2, 9) | -6.9584 | -7.8672 | -8.4093 ($A_2$) | -8.1914 | |
| Coronene | a (4, 24) | -10.2764 | -11.2733 | -11.7966 (A) | -11.4628 | -11.9755 |
| | b (4, 18) | -9.7660 | -10.7447 | -11.6399 ($A_1$) | -11.3781 | |
| | c (6, 21) | -10.5736 | -11.3457 | -11.8103 ($A_1$) | -11.3740 | |
| | d (8, 24) | **-10.8304** | **-11.6961** | **-11.9459** ($A_1$) | **-11.4992** | |
| | e (2, 12) | -9.5702 | -10.6676 | -11.2997 ($A_1$) | -11.3109 | |
| | f (2, 12) | -9.6055 | -10.7084 | -11.6190 ($A_1$) | -11.3968 | |
| | g (6, 18) | -9.9867 | -10.9182 | -11.6693 ($A_1$) | -11.1837 | |
| | h (6, 12) | -9.4117 | -11.1230 | -11.8635 ($A_1$) | -11.1287 | |
| Hexabenzo-coronene | a (6, 42) | **-19.8044** | -20.4297 | -21.0044 ($B_1$) | -20.7082 | _[b] |
| | b (7, 42) | -19.6337 | **-20.4824** | **-21.0786** ($B_1$) | **-20.7724** | |
| Hexa-*cata*-hexabenzo-coronene | a (6, 42) | -21.5999 | -23.0983 | **-23.8296** ($A_1$) | -23.1753 | _[b] |
| | b (8, 48) | **-22.4271** | **-23.3169** | -23.7980 (A) | **-23.4848** | |
| | c (8, 42) | -21.1057 | -21.9885 | -23.0109 ($A_1$) | -23.0836 | |
| Kekulene | a (6, 48) | -21.4983 | -22.2249 | -22.7387 ($A_1$) | -22.9403 | _[b] |
| | b (6, 36) | -19.8349 | -20.9791 | -22.4499 ($A_1$) | -22.2943 | |
| | c (8, 48) | -21.4735 | -22.2083 | -23.0747 ($A_1$) | -22.9187 | |
| | d (8, 48) | **-22.1306** | **-22.7459** | **-23.3603** (A) | **-23.2404** | |

[a]Labels refer to Figure 6–Figure 10; cluster size $q$ and the number of intracluster bonds are given in parentheses. [b]System size exceeds our ED capabilities.

In triphenylene, grouping (b) is obtained from (a) by merging clusters, while conserving the full PG-symmetry, and (b) is thus guaranteed to yield lower cPHF energies than (a). The same is true for (c) with respect to the two complementary dimer formations (d) and (e). In fact, $D_3$SGHF(6) based on (c) with three benzene units, each hosting a Clar sextet in the most favorable resonance structure [32], is very close to the exact ground state.

In coronene, grouping (d), including 24 of 30 bonds in the clusters, is obtained from (b) by merging neighboring clusters pairwise and recovers 99.75% of $E_0$ with $D_3$SGHF(8). For cluster types (a)–(d), Table 3 collects all distinct nearest-neighbor (NN) and a few next-nearest neighbor (NNN) SPCFs. Their magnitude is generally somewhat over- or underestimated (with minor exceptions), depending on whether sites reside in the same (bold type in Table 3) or in different clusters, respectively. It thus appears generally advantageous to include pairs of strongly antiferromagnetically correlated sites in a cluster. The SPCFs evidence a moderate degree of spatial-symmetry breaking, $D_6 \to C_6$ in (a), $D_6 \to D_3$ in (c), and $D_6 \to D_3$ in (d), but not in (b), because this grouping allows full $D_6$-projection.

Table 3: SPCFs $\langle \hat{\mathbf{s}}_i \cdot \hat{\mathbf{s}}_j \rangle$ in coronene from PGSGHF, compared against exact results. Letters (a), (b), (c), (d) identifying cluster groupings, and the site numbers (first column) are defined in Figure 7.[a]

| $i$–$j$ | $C_6$SGHF(4) (a) | $D_6$SGHF(4) (b) | $D_3$SGHF(6) (c) | $D_3$SGHF(8) (d) | Exact |
|---|---|---|---|---|---|
| 1–2 | -0.33729 | -0.35059 | **-0.39063** | **-0.36850** | -0.35875 |
| 2–3 | -0.33729 | -0.35059 | **-0.38706** | -0.33627 | -0.35875 |
| 1–4 | **-0.40404** | **-0.35322** | -0.30621 | **-0.37926** | -0.37507 |
| 4–5 | **-0.37034** | **-0.40476** | **-0.37196** | **-0.37095** | -0.36665 |
| 5–6 | **-0.54911** | -0.42665 | **-0.52422** | **-0.52685** | -0.52881 |
| 6–7 | -0.30533 | **-0.40476** | **-0.37196** | **-0.37095** | -0.36665 |
| 7–8 | **-0.37034** | **-0.40476** | **-0.41287** | **-0.36483** | -0.36665 |
| 8–9 | **-0.54911** | -0.42665 | -0.45278 | -0.52027 | -0.52881 |
| (1–3) | 0.17744 | 0.19472 | **0.19503** | 0.16792 | 0.16641 |
| (1–5) | **0.18265** | **0.20098** | 0.16410 | **0.17971** | 0.17661 |
| (2–6) | 0.18009 | **0.20098** | 0.16410 | **0.17971** | 0.17661 |
| (4–6) | **0.20777** | 0.18967 | **0.19643** | 0.19440 | 0.19291 |
| (5–7) | 0.17515 | 0.18967 | **0.19643** | 0.19440 | 0.19291 |

[a]SPCFs for all distinct NN pairs $i$–$j$ (first column) are given, including pairs that are equivalent in the full symmetry group, but inequivalent in some of the PGSGHF wave functions. The NNN set (pairs $i$–$j$ in parentheses) is not complete, except for (b), which maintains the full $D_6$ symmetry. Bold type is used for SPCFs of pairs that belong to the same cluster.

Due to computational limitations, we cannot carry out ED on the $N = 42$ and $N = 48$ lattices, but a comparison with cPT2 suggests that cPHF is still fairly accurate in these larger systems. In hexabenzocoronene, both options include a fraction of 42/54 bonds in the clusters. Although $q = 6$ (Figure 8a) has seven intact rings, and thus loosely corresponds to the most favorable resonance structure with seven Clar sextets [32], $q = 7$ (Figure 8b) with six sextets is energetically very slightly favored in $D_6$SGHF, though not in UHF (Table 2). In hexa-*cata*-hexabenzocoronene, $D_6$SGHF(6) with 42/60 bonds in seven rings (Figure 9a) yields a significantly lower energy than $D_6$SGHF(8) with the same number of intracluster bonds but no intact rings (Figure 9c), and is even slightly better than $C_6$SGHF(8) with 48 intracluster bonds, but only six rings (Figure 9b). Finally, in kekulene, grouping (d) predicts the lowest energy, despite $D_6 \rightarrow C_6$ symmetry breaking. In contrast to (a), (c) and (d), the six Clar sextets are broken up in (b), which features only 36/60 bonds, and yields higher energies in cHF, cPHF and cPT2.

**Polyhedra.** We lastly consider four polyhedra with $T_d$ or $I_h$ symmetry, which allow cluster groupings that fully respect spatial symmetry.

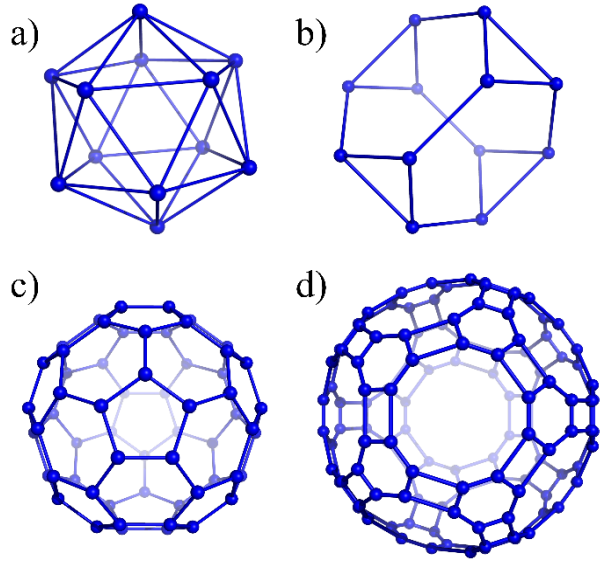

Figure 11: Icosahedron (a), truncated tetrahedron (b), truncated icosahedron (c), and truncated icosidodecahedron (d).

a) Icosahedron

Dimers of the sites related by spatial inversion $C_i$ conserve $I_h$ symmetry (Figure 12a), while nearest-neighbor pairs break symmetry ($D_{2h}$, Figure 12b), as do hexamers ($D_{5d}$, Figure 12c). For $s > \frac{1}{2}$, the antiferromagnetic spin-pair correlation $\langle \hat{\mathbf{s}}_1 \cdot \hat{\mathbf{s}}_4 \rangle$ between the most distant sites (numbers defined in Figure 12a) exceeds the NN correlation $\langle \hat{\mathbf{s}}_1 \cdot \hat{\mathbf{s}}_2 \rangle$ in the exact wave functions, see Table 5.

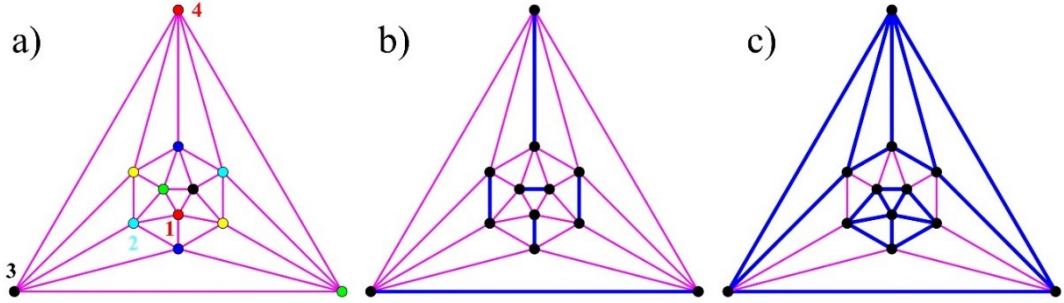

Figure 12: Three cluster groupings in the icosahedron conserve $I_h$ (a, $q = 2$), $D_{2h}$ (b, $q = 2$) or $D_{5d}$ symmetry (c, $q = 6$). In the planar projections (Schlegel diagrams), interacting sites belonging to the same or to different clusters are connected by blue or pink lines, respectively. In (a), the color of sites assigns them to one of six clusters. Sites forming symmetry-inequivalent pairs with site 1 are numbered in (a).

Interestingly, the $I_h$ grouping (a) is advantageous over $D_{2h}$ (b), see Table 4, except for GHF, where (a) yields the classical solution [33] while (b) profits energetically from correlation in the dimers. For $s = \frac{1}{2}$, GHF yields local singlets in (b), which are broken up in the SGHF reference, as apparent from the energy lowering, $E_{\text{SGHF}} < E_{\text{GHF}}$.

Table 4: GHF and PHF estimates of ground-state energies of the antiferromagnetic icosahedron with $\frac{1}{2} \leq s \leq 2$ for two different $q = 2$ groupings (Figure 12a and b) with $I_h$ or $D_{2h}$ symmetry.

| $s$ | Grouping | GHF | SGHF | PGSGHF | Exact (Γ) |
|---|---|---|---|---|---|
| 1/2 | $D_{2h}$ | -4.5000 | -5.3224 | -6.1717 | -6.1879 |
| | $I_h$ | -3.3541 | -5.7644 | -6.1879[a] | ($A_u$) |
| 1 | $D_{2h}$ | -14.3025 | -17.4565 | -18.1678 | -18.5611 |
| | $I_h$ | -13.4164 | -18.2225 | -18.5609 | ($A_g$) |
| 3/2 | $D_{2h}$ | -31.4256 | -36.2633 | -37.3073 | -37.7412 |
| | $I_h$ | -30.1869 | -37.3842 | -37.7396 | ($A_u$) |
| 2 | $D_{2h}$ | -55.2658 | -61.7751 | -63.1481 | -63.7104 |
| | $I_h$ | -53.6656 | -63.2529 | -63.7075 | ($A_g$) |

[a]Exact ground-state energy within numerical double precision.

For $s = \frac{1}{2}, 1, \frac{3}{2}, 2$, $I_h$SGHF(2) yields 100%, 99.999%, 99.996% and 99.995%, respectively, of the exact ground-state energy and thereby recovers most of the correlation energy missing from $I_h$SGHF(1) [6]. Not surprisingly, SPCFs are very close to the exact results (Table 5), corroborating the high quality of the $I_h$SGHF(2) wave functions.

Table 5: Comparison of $I_h$SGHF(2) predictions of SPCFs in the ground state of the icosahedron with $\frac{1}{2} \leq s \leq 2$ against exact values. Site numbers are defined in Figure 12a.

| $s$ | | $\langle \hat{\mathbf{s}}_1 \cdot \hat{\mathbf{s}}_2 \rangle$ | $\langle \hat{\mathbf{s}}_1 \cdot \hat{\mathbf{s}}_3 \rangle$ | $\langle \hat{\mathbf{s}}_1 \cdot \hat{\mathbf{s}}_4 \rangle$ |
|---|---|---|---|---|
| 1/2 | Exact | -0.2063 | 0.0841 | -0.1397 |
| | PHF | -0.2063 | 0.0841 | -0.1397 |
| 1 | Exact | -0.6187 | 0.3680 | -0.7463 |
| | PHF | -0.6187 | 0.3680 | -0.7464 |
| 3/2 | Exact | -1.2580 | 0.9060 | -1.9899 |
| | PHF | -1.2580 | 0.9062 | -1.9910 |
| 2 | Exact | -2.1237 | 1.6616 | -3.6897 |
| | PHF | -2.1236 | 1.6621 | -3.6926 |

Incidentally, $I_h$GHF(2) (without explicit S-projection) converges onto the exact ground state of the $s = \frac{1}{2}$ system (the numerical deviation from $E_0$ is $\approx 10^{-15}$). In contrast, in $I_h$KGHF(1) the additional use of complex-conjugation symmetry (K) was required to reach the exact solution [6]. The fact that the exact ground state for $s = \frac{1}{2}$ is also obtained with $D_{5d}$SGHF(6) (Figure 12c) shows that a cluster grouping that breaks symmetry does not necessarily lead to a symmetry-broken cPHF wave function. SGHF(2) wave functions are totally symmetric in group $I$, but have mixed symmetry under spatial inversion $C_i$ ($I_h = I \otimes C_i$). For $s = \frac{1}{2}$, the respective weights in the SGHF(2) wave function are $w_{A_g} \approx 0.418$ and $w_{A_u} \approx 0.582$ ($w_{A_g} + w_{A_u} = 1$).

b) Truncated Tetrahedron

The classical solution for the truncated tetrahedron [34] minimizes frustration on the four triangles (with angles of 120° between spins) and aligns spins on the bonds between triangles antiparallel. Two complementary cluster groupings defined by either six classically unfrustrated bonds (u-bonds, $q = 2$, Figure 13a) or twelve frustrated bonds (f-bonds, $q = 3$, Figure 13b) maintain tetrahedral $T_d$ symmetry.

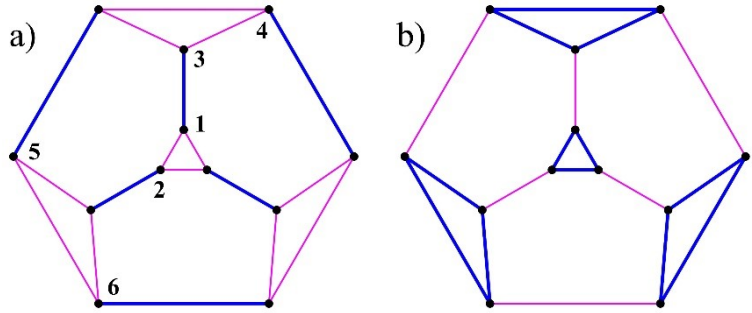

Figure 13: Dimers (a) and trimers (b) respect the full $T_d$ symmetry of the truncated tetrahedron. For more details, see caption to Figure 12.

Table 6: GHF and PHF estimates of ground-state energies of the antiferromagnetic $\frac{1}{2} \leq s \leq 2$ truncated tetrahedron for two different groupings (Figure 13).

| $s$ | $q$ | GHF | SGHF | $T_d$SGHF | Exact ($\Gamma$) |
|---|---|---|---|---|---|
| 1/2 | 2 | -4.5000 | -5.2700 | -5.7009[a] | -5.7009 |
|     | 3 | -3.8881 | -4.8147 | -5.7009[a] | (A$_2$) |
| 1   | 2 | -14.0173 | -16.0342 | -17.1649 | -17.1955 |
|     | 3 | -13.8696 | -15.7195 | -17.1775 | (A$_1$) |
| 3/2 | 2 | -29.7756 | -32.8938 | -34.4456 | -34.6402 |
|     | 3 | -29.6977 | -32.5614 | -34.4796 | (A$_2$) |
| 2   | 2 | -51.5616 | -55.7815 | -57.7827 | -58.1140 |
|     | 3 | -51.5327 | -55.3924 | -57.8181 | (A$_1$) |

[a]Exact ground-state energy within numerical double precision.

Table 6 shows that u-bonds ($q = 2$) yield lower GHF and SGHF energies than f-bonds ($q = 3$), despite a smaller number of variational parameters, e.g., for $s = 2$, $N_{var} = 288$ for $q = 2$, and $N_{var} = 992$ for $q = 3$. However, the larger variational freedom provided by f-bonds leads to slightly better energies for $T_d$SGHF(3) compared to $T_d$SGHF(2). The somewhat larger deviations between PGSGHF and exact results than in the icosahedron may be attributed to the lower order of the group ($h = 24$ in $T_d$ versus $h = 120$ in $I_h$), although a direct comparison between these two polyhedra is not meaningful. For $q = 1$ [6] and $q = 2$, though not for $q = 3$, and for all tested values of $s$, the SGHF reference $|\Phi\rangle$ can be revealed to have the classical spin-density structure by appropriate gauge transformations [21]. (For additional comments on this issue, see the following section on the truncated icosahedron.) SGHF(1) [6] and SGHF(2) wave functions transform like the exact ground state ($A_2$ or $A_1$ for half-integer or integer $s$, respectively).

c) Truncated Icosahedron

In contrast to icosidodecahedral arrangements of spin centers in Keplerate molecules $\{Mo_{72}V_{30}\}$ [35,36], $\{W_{72}V_{30}\}$ [37], $\{Mo_{72}Cr_{30}\}$ [38], $\{Mo_{72}Fe_{30}\}$ [39,40], with $s = \frac{1}{2}, \frac{3}{2}, \frac{5}{2}$ for $V^{3+}$, $Cr^{3+}$, $Fe^{3+}$, truncated icosahedra were not yet synthetically realized as magnetic molecules. The respective Heisenberg model was nevertheless addressed in a few works [27,41–47], some of which were motivated by gaining an understanding of the properties of buckminsterfullerene $C_{60}$. However, electronic-structure calculations have shown that the Heisenberg model is at best of qualitative value for $C_{60}$, because this molecule possesses a number of unpaired electrons $\ll 60$ [48] (but note that there is no unique measure for this number). It was eventually concluded that $C_{60}$ displays no significant strong correlation [49]. *Ab initio* GHF calculations [48] on $C_{60}$ still reproduce the exotic three-dimensional spin-density pattern of the classical Heisenberg ground state [27]. Geometry optimization on the GHF level-of-theory yields a perfect $I_h$ structure for $C_{60}$ [48], but the spatial inversion $C_i$ is the only obvious self-consistent symmetry of the GHF solution. We specify the hidden icosahedral symmetry $I$ in Figure 14, where classical spin vectors are plotted in a Schlegel diagram. A uniform rotation was applied such that spins in the central pentagon in Figure 14 lie in the geometrical plane of that pentagon, with the spin on the first site pointing in the negative *x*-direction. This spin configuration is left unchanged by a combined spin rotation (R) by 144° and a five-fold site permutation (P). In the electronic-structure problem, the site permutation corresponds to a spatial rotation. These operations are performed about an axis through the coordinate origin (the center of the truncated icosahedron) and the center of the central pentagon (marked $5_{PR}$ in Figure 14).

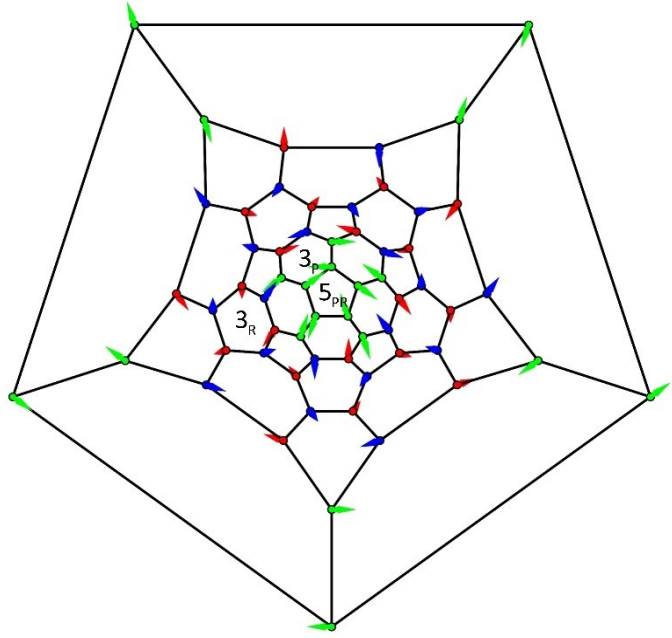

Figure 14: Classical three-dimensional spin configuration of the truncated icosahedron. Green spin vectors lie in the $xy$-plane (paper plane), red/blue vectors point in the negative/positive $z$-direction. The first vector on the central pentagon points in the (horizontal) negative $x$-direction. Symmetry elements ($3_P \times 3_R = 3_{PR}$ and $5_{PR}$, see main text for details) are defined with respect to an axis through the coordinate origin and the center of the respective pentagon or hexagon.

A spin rotation by 120° about an axis through the mid-point of a hexagon ($3_R$, Figure 14), combined with a threefold site permutation about another hexagon ($3_P$) is a second symmetry ($3_P \times 3_R = 3_{PR}$), and $5_{PR}$ and $3_{PR}$ are generators of a group isomorphic to group $I$. The full symmetry group of the classical (or GHF) solution is $I_h = I \otimes C_i$. This overall explains why GHF orbitals of $C_{60}$ are up to six-fold degenerate: molecular orbitals span irreducible representations [50] of the double group $I_h^*$.

Two complementary cluster formations preserve spatial $I_h$ symmetry by including either the 30 u-bonds between pentagons ($q = 2$, Figure 15a) or the 60 f-bonds in the pentagons ($q = 5$, Figure 15b). Classical spins are antiparallel on u-bonds (see Figure 14) and span angles of 144° on the f-bonds. We additionally consider fused hexagons ($q = 10$, Figure 15c), thereby including 36 f-bonds and all 30 u-bonds in the clusters and reducing symmetry, $I_h \to D_{2h}$.

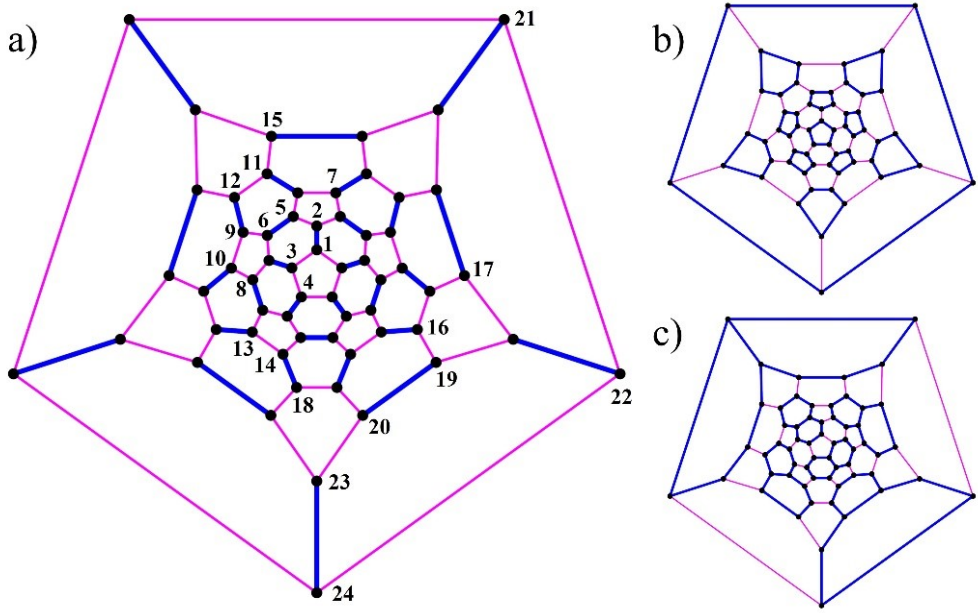

Figure 15: Inter-pentagon bonds ($q = 2$, a) or pentagon clusters ($q = 5$, b) maintain the $I_h$ symmetry of the truncated icosahedron. Fused hexagons ($q = 10$, c) reduce symmetry to $D_{2h}$. The numbering (a) of centers forming inequivalent pairs with site 1 follows Fig. 3 in Ref. [41].

Interestingly, GHF(2) can be solved analytically for $s = \frac{1}{2}$. Starting from the classical solution [27], $E_{HF} = -\frac{15}{4}(3+\sqrt{5}) \approx -19.6353$, we define a u-bond wave function $|\Phi_i\rangle$ in Eq. (14),

$$|\Phi_i\rangle = \frac{1}{\sqrt{1+\alpha^2}}\left(|\uparrow\downarrow\rangle + \alpha |\downarrow\uparrow\rangle\right), \qquad (14)$$

that depends on a real parameter $\alpha$. The local quantization axis for projections $\uparrow$ and $\downarrow$ in a cluster $i$ are given by the classical solution, that is, $\alpha = 0$ recovers GHF(1). Thus, all spins retain their classical orientation in GHF(2), but Eq. (14) allows the local singlet to acquire a higher weight than the triplet. Each dimer contributes an energy $\lambda$, Eq. (15):

$$\lambda = \langle \Phi_i | \hat{\mathbf{s}}_1 \cdot \hat{\mathbf{s}}_2 | \Phi_i \rangle = \frac{1}{1+\alpha^2}\left(-\frac{\alpha^2}{4} + \alpha - \frac{1}{4}\right). \qquad (15)$$

Compared to GHF(1), the local magnetization $|\langle \hat{\mathbf{s}}_m \rangle|$ is reduced by a factor $\kappa$, Eq. (16), in the GHF(2) wave function.

$$\kappa = \frac{1-\alpha^2}{1+\alpha^2} \qquad (16)$$

Interactions along the intercluster bonds (f-bonds) are evaluated classically,

$$\langle \hat{\mathbf{s}}_m \cdot \hat{\mathbf{s}}_n \rangle = \langle \hat{\mathbf{s}}_m \rangle \cdot \langle \hat{\mathbf{s}}_n \rangle = \left(\frac{\kappa}{2}\right)^2 \cos\phi , \tag{17}$$

with $\phi = \frac{4\pi}{5}$ (144°). Minimization of the total GHF(2) energy $E$, Eq. (18),

$$E = 30\lambda + 60\left(\frac{\kappa}{2}\right)^2 \cos\left(\frac{4\pi}{5}\right), \tag{18}$$

with respect to $\alpha$ affords $\alpha_{opt}$ in Eq. (19),

$$\alpha_{opt} = -\frac{1}{4}\left[\left(1+\sqrt{5}\right)\left(2+\sqrt{2}\sqrt{\sqrt{5}-1}\right)\right], \tag{19}$$

corresponding to $\kappa_{opt}$ in Eq. (20),

$$\kappa_{opt} = \left|\frac{1-\alpha_{opt}^2}{1+\alpha_{opt}^2}\right| = \sqrt{\frac{\sqrt{5}-1}{2}}, \tag{20}$$

and overall results in Eq. (21),

$$E_{GHF(2)} = -\frac{15}{2}\left(1+\sqrt{5}\right) \approx -24.2705. \tag{21}$$

Numerical calculations indeed converge onto this solution. (An analogous analysis for the $s = \frac{1}{2}$ truncated tetrahedron yields $\alpha_{opt} = 1$, that is, singlets are formed on the six u-bonds.) GHF(2) significantly improves over GHF(1), but still falls short of an accurate description. Other works derived the following variational estimates for the $S = 0$ ground state: $E = -29.97$ [43] (resonating valence-bond, RVB), $E = -30.83$ [42] or $E = -30.69$ [41] (Variational Monte Carlo with a Gutzwiller projector, VMC), and $E = -31.13$ [45] (DMRG). The latter most recent result (DMRG) may be regarded as a quasi-exact benchmark value. The GHF and PHF energies ($q = 2, 5, 10$) are collected in Table 7. Our best estimate of $E = -29.98$, obtained with $D_{2h}$SGHF(10), is very close to the RVB result [43] and corresponds to $p = 90\%$. In view of the Hilbert-space dimension of $\mathcal{N} = 2^{60} \approx 10^{18}$, which is presently still infeasible for ED, even if symmetries were employed to reduce the matrix size, we would like to emphasize the very significant state-space reduction afforded by cPHF in terms of $N_{var} = 180$ for $q = 2$, $N_{var} = 744$ for $q = 5$, and $N_{var} = 12276$ for $q = 10$.

Table 7: cGHF and cPHF estimates of ground-state energies of the $s = \tfrac{1}{2}$ truncated icosahedron.

| $q$ | GHF | SGHF | PGSGHF[a] |
|---|---|---|---|
| 2 | -24.2705 | -25.5486 | -27.8429 |
| 5 | -25.8525 | -26.6072 | -28.5653 |
| 10 | -28.6199[b] | -29.2195 | -29.9842 |

[a]Projection onto $S = 0$, $\Gamma = A_g$. [b]All clusters assume their local singlet ground state.

As explained, GHF(2) spin densities (local magnetizations) maintain the classical structure. Our numerical calculations showed that this also holds true for GHF(5). Although SGHF(2) or SGHF(5) will generally converge onto a reference $|\Phi\rangle$ that does not display the classical spin-density pattern, we found that the latter can be restituted by gauge rotations on $|\Phi\rangle$ that leave the S-projected state $\hat{P}_0^0|\Phi\rangle$ unchanged (redundancies in the definition of $|\Phi\rangle$ with respect to gauge transformations related to spin symmetry were discussed in Ref. [21]): $|\Phi\rangle$ assumes the classical magnetization pattern when it is ensured through appropriate gauge transformations that the expectation value of the total-spin vector vanishes, $\langle\Phi|\hat{\mathbf{S}}|\Phi\rangle = \mathbf{0}$. Irrespective of whether such transformations are applied to $|\Phi\rangle$, SGHF(2) and SGHF(5) wave functions ($S = 0$) have pure $A_g$ symmetry in $I_h$. For $q = 1$, $|\Phi_{HF}\rangle$ is the optimal reference for $S = 0$ projection in SGHF ($E_{SGHF(1)} = -21.6515$), which appears to be a rather general feature of highly symmetric polyhedra [6], but is neither true for spin rings [6], nor for the presently studied polyhedra if $q > 1$. In other words, for $q > 1$, the variation-after-projection (VAP) approach of SGHF is not equivalent to a far simpler projection-after-variation (PAV) "single-shot" S-projection of $|\Phi_{HF}\rangle$.

In Table 8, PHF predictions for SPCFs from SGHF and $I_h$SGHF ($q = 2, 5$) are compared to VMC [41]. Note that the energy of a variational trial function that respects icosahedral symmetry is $E = 30\langle\hat{\mathbf{s}}_1\cdot\hat{\mathbf{s}}_2\rangle + 60\langle\hat{\mathbf{s}}_1\cdot\hat{\mathbf{s}}_3\rangle$. The magnitude of the antiferromagnetic u-bond correlation $\langle\hat{\mathbf{s}}_1\cdot\hat{\mathbf{s}}_2\rangle$ is underestimated by $I_h$SGHF(5), while the magnitude of $\langle\hat{\mathbf{s}}_1\cdot\hat{\mathbf{s}}_3\rangle$ in the f-bonds is overestimated. The reverse is true for $I_h$SGHF(2), which yields $\langle\hat{\mathbf{s}}_1\cdot\hat{\mathbf{s}}_2\rangle$ and $\langle\hat{\mathbf{s}}_1\cdot\hat{\mathbf{s}}_3\rangle$ values that match the VMC estimate more closely. This further illustrates the trend observed in the hexagonal lattices of over- or underestimating correlations in intra- or intercluster bonds, respectively. Except for $I_h$SGHF(2), cPHF overestimates long-range correlation rather dramatically.

Table 8: SPCFs, $\langle \hat{\mathbf{s}}_1 \cdot \hat{\mathbf{s}}_j \rangle$, for all distinct pair types (numbering defined in Figure 15) in the ground state of the truncated icosahedron. VMC results were taken from Table II in Ref. [41].

|   | $q = 2$ | | $q = 5$ | | |
|---|---|---|---|---|---|
| $E$ | -25.5486 | -27.8429 | -26.6072 | -28.5653 | -30.69 |
| $j$ | SGHF | $I_h$SGHF | SGHF | $I_h$SGHF | VMC |
| 2 | -0.562 | -0.610 | -0.186 | -0.277 | -0.529 |
| 3 | -0.145 | -0.159 | -0.351 | -0.337 | -0.247 |
| 4 | 0.051 | 0.051 | 0.076 | 0.073 | 0.030 |
| 5 | 0.136 | 0.137 | 0.142 | 0.154 | 0.141 |
| 6 | -0.145 | -0.154 | -0.151 | -0.154 | -0.142 |
| 7 | -0.056 | -0.054 | -0.059 | -0.061 | -0.023 |
| 8 | -0.090 | -0.080 | -0.094 | -0.093 | -0.038 |
| 9 | 0.084 | 0.070 | 0.087 | 0.083 | 0.031 |
| 10 | -0.002 | 0.001 | -0.003 | -0.001 | 0.001 |
| 11 | 0.051 | 0.049 | 0.052 | 0.051 | 0.027 |
| 12 | -0.090 | -0.072 | -0.094 | -0.088 | -0.026 |
| 13 | -0.090 | -0.042 | -0.094 | -0.084 | -0.002 |
| 14 | 0.051 | 0.017 | 0.052 | 0.046 | -0.001 |
| 15 | -0.002 | -0.002 | -0.003 | -0.003 | -0.004 |
| 16 | 0.084 | 0.037 | 0.087 | 0.078 | 0.001 |
| 17 | -0.090 | -0.036 | -0.094 | -0.081 | 0.002 |
| 18 | -0.056 | -0.018 | -0.059 | -0.051 | 0.013 |
| 19 | -0.145 | -0.042 | -0.151 | -0.129 | 0.000 |
| 20 | 0.136 | 0.039 | 0.142 | 0.124 | -0.002 |
| 21 | 0.051 | 0.016 | 0.053 | 0.046 | -0.030 |
| 22 | -0.145 | -0.040 | -0.150 | -0.128 | 0.007 |
| 23 | -0.179 | -0.046 | -0.186 | -0.158 | 0.016 |
| 24 | 0.168 | 0.044 | 0.176 | 0.152 | -0.008 |

We are not aware of any previous works addressing ground states in the truncated icosahedron with $s > \frac{1}{2}$. For $\frac{1}{2} \leq s \leq 2$, energies from GHF, SGHF and $I_h$SGHF ($q = 2$) are collected in Table 9, where they are compared with local singlets on fused benzene rings ($q = 10$), and with PT2(1) and PT2(2). Except for $s = \frac{1}{2}$, where the singlet-product represents the GHF(10) solution, $I_h$SGHF yields the best variational energy. For the $s = \frac{1}{2}$ system, the PT2(1) energy, $E = -31.05$ [27], is incidentally rather close to the DMRG result, $E = -31.13$ [45], but the PT2(2) prediction deviates significantly, $E = -29.12$.

Table 9: Variational estimates from GHF, SGHF, $I_h$SGHF (projection onto $S = 0$, $\Gamma = A_g$), a singlet-product on fused hexagons $(q = 10)$, and PT2 for the ground state of the truncated icosahedron with $\tfrac{1}{2} \leq s \leq 2$.

| $s$ | GHF(2) | SGHF(2) | $I_h$SGHF(2) | $q = 10$, $s_i = 0$ | PT2(1) | PT2(2) |
|---|---|---|---|---|---|---|
| 1/2 | -24.2705 | -25.5486 | -27.8429 | -28.6199 | -31.0543 | -29.1216 |
| 1 | -85.6371 | -87.7764 | -90.1147 | -89.8943 | -96.7113 | -96.9910 |
| 3/2 | -186.6961 | -189.5706 | -192.4293 | -183.7630 | -202.2428 | -203.7112 |
| 2 | -327.0802 | -330.4438 | -343.1173 | -310.5941 | -347.1526 | -349.5852 |

The relative deviation between PT2 energies and $I_h$SGHF is smallest for $s = 2$. The Hilbert-space of dimension $\mathcal{N} = 5^{60} \approx 9 \times 10^{41}$ for $s = 2$ is completely out of reach of ED. The problem size is again drastically reduced by cPHF(2), which operates with merely $N_{\text{var}} = 1440$ variational parameters. Somewhat unfortunately, we cannot offer a reliable assessment of the accuracy of cPHF in this rather large system, where obtaining an accurate reference value with DMRG would be very challenging.

d) Truncated icosidodecahedron

Lastly, as another demonstration of the applicability of cPHF to systems that are far too large for ED, we consider a truncated icosidodecahedron with 120 $s = \tfrac{1}{2}$ vertices, $\mathcal{N} = 2^{120} \approx 10^{36}$. Figure 16 presents three symmetry-compatible cluster groupings $(q = 4, 6, 10)$.

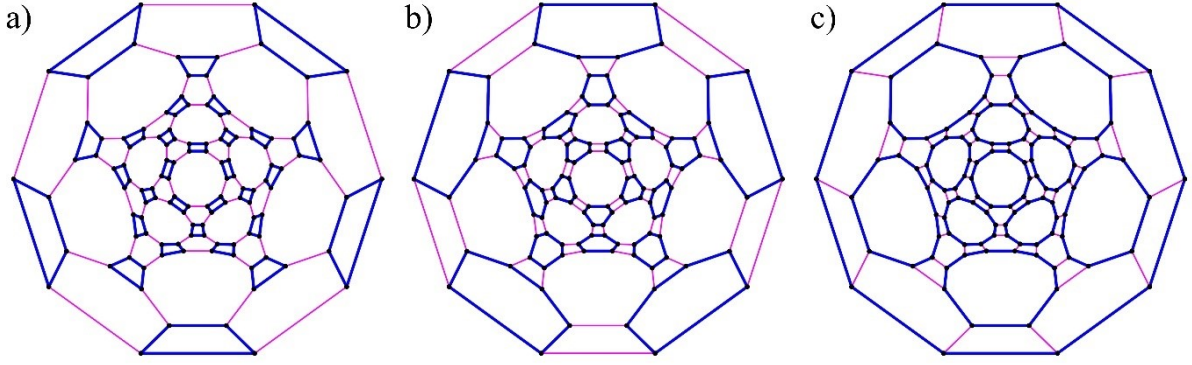

Figure 16: Cluster groupings compatible with $I_h$ symmetry in the truncated icosidodecahedron. The midpoints of the squares ($q = 4$, a), hexagons ($q = 6$, b) and decagons ($q = 10$, c) form an icosidodecahedron ($Q = 30$), dodecahedron ($Q = 20$) or icosahedron ($Q = 12$), respectively.

Being bipartite, the system has a non-degenerate $S = 0$ ground state, which should transform as $A_g$ (like the classical Néel state, $E_{UHF(1)} = -45$). With 120/180 bonds included in the clusters, none of the three groupings has an obvious advantage: $E_{UHF(4)} = -60.7642$, $E_{UHF(6)} = -59.0690$, $E_{UHF(10)} = -58.8657$. For a given $q$, all clusters assume the same $m_i = 0$ state, but no local singlets are formed ($s_i \neq 0$), leading to spin densities of Néel-type. We selected $q = 4$ for cPHF calculations ($N_{var} = 900$): $E_{SUHF} = -61.3769$, $E_{SGHF} = -61.6502$, $E_{I_hSGHF} = -64.0101$. With respect to $I_h$SGHF, $q = 6$ ($N_{var} = 2520$) is still inferior to $q = 4$, yielding $E_{I_hSGHF} = -62.7728$. The same trend is true for PT2: $E_{PT2(4)} = -64.9139$, $E_{PT2(6)} = -64.0437$.

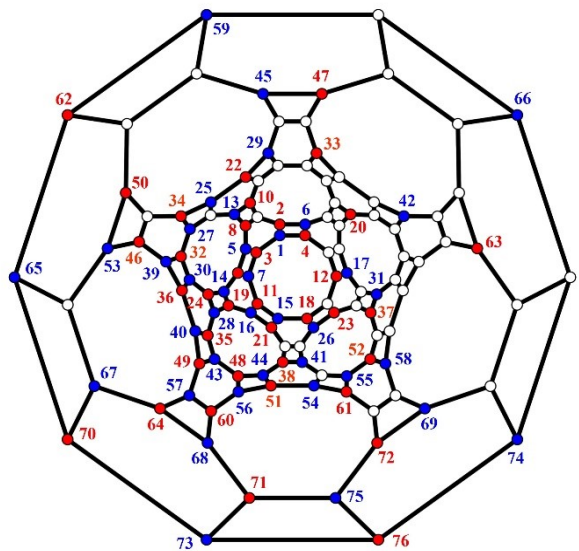

Figure 17: Numbering of centers forming inequivalent pairs with site 1 in the bicolorable (red, blue) truncated icosidodecahedron lattice. Sites without a number are white, but still belong to one of the two sublattices.

SPCFs from $I_h$SGHF(4) are plotted in Figure 18. All correlations within the same sublattice are positive, and negative between different sublattices. The strong correlation across the whole range of the molecule is most likely artifactual, because PHF (or cPHF) reverts to HF (or cHF) in the thermodynamic limit [5,6]. Therefore, long-range order for large systems is generally exaggerated.

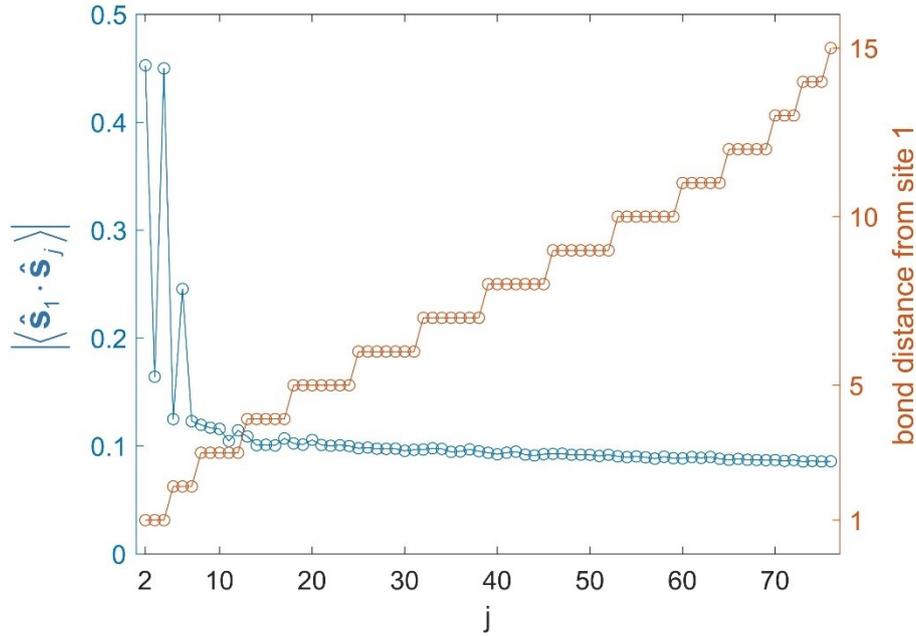

Figure 18: Magnitude of SPCFs (left *y*-axis; correlations are positive/negative for pairs in same/different lattices) and bond distances from reference site 1 (right *y*-axis) in the truncated icosidodecahedron. The site numbering follows Figure 17.

## 4. Conclusions

By partitioning spin sites into clusters, the cPHF method extends the variational flexibility of PHF for a simple approximation of ground states of finite Heisenberg systems. The optimization of a cluster-product state for the restoration of good quantum numbers (spin and point group) comes at a mean-field computational cost, with a prefactor depending on the projection-grid size, and the compact representation of the cPHF wave function in terms of a projector acting on a cluster mean-field state is suitable for the calculation of various properties. We considered only energies and spin-pair correlations, but other quantities needed for modeling EPR or INS spectra, such as spin densities or expectation values of higher-rank local spin operators, as well as transition-density matrices, could also be obtained straightforwardly, opening a perspective for the application of cPHF to moderately large magnetic molecules.

For antiferromagnetic $s = \tfrac{1}{2}$ spin rings, which are more challenging for PHF than $s > \tfrac{1}{2}$ systems [6], cPHF significantly improves over ordinary PHF by predicting rather accurate ground states for larger ring sizes. Although the cluster ansatz cannot access the full cyclic symmetry, the accuracy of cPHF improves with cluster size, where a smaller number of bonds is left to be correlated through symmetry projection.

We additionally studied hexagonal lattice fragments ($s = \frac{1}{2}$) and symmetric polyhedra ($\frac{1}{2} \leq s \leq 2$) where cluster groupings can maintain the full spatial symmetry. In these systems, it is generally advantageous to include strongly antiferromagnetic bonds in the clusters. This may be accomplished by maximizing the number of intact rings (Clar sextets) in hexagonal lattices or by defining, if possible, clusters in terms of classically unfrustrated bonds in polyhedra. Applications to the $s = 2$ truncated icosahedron and the $s = \frac{1}{2}$ truncated icosidodecahedron demonstrate that cPHF can be applied to systems whose size is prohibitive for exact diagonalization and challenging for other methods.

More advanced symmetry-projected methods have been under active development in various fields of many-body physics and appear worth pursuing for Heisenberg systems. Specifically, a linear combination of cluster-product states (configurations) is systematically improvable by increasing the number of configurations and would thus ameliorate problems associated with the lack of size-extensivity. Such a multi-configuration variant of cPHF would also give access to excited states in the respective symmetry sectors that are needed for modeling spectra and low-temperature properties of molecular magnets.

**Acknowledgement.** SGT thanks the German Academic Exchange Service (DAAD) for support in the early stages of this project.

# Ground states of Heisenberg spin clusters from a cluster-based projected Hartree-Fock approach

Shadan Ghassemi Tabrizi[1,a,*] and Carlos A. Jiménez-Hoyos[1,†]

[1]*Department of Chemistry, Wesleyan University, Middletown, CT 06459, USA*

[a]*Present address: Department of Chemistry, University of Potsdam, Karl-Liebknecht-Str. 24-25, D-14476, Potsdam-Golm, Germany*

*\*shadan_ghassemi@yahoo.com*

[†]*cjimenezhoyo@wesleyan.edu*

## Supporting Information

### 1. Optimization of the broken-symmetry reference

The self-consistent field (SCF) optimization of the cluster-product state $|\Phi\rangle$, Eq. (S1),

$$|\Phi\rangle = |\Phi_1\rangle|\Phi_2\rangle\ldots|\Phi_Q\rangle = \prod_{i=1}^{Q}|\Phi_i\rangle \quad , \tag{S1}$$

in cPHF is closely analogous to the diagonalization-based PHF algorithm described in the Supplemental Material to Ref. [S1]. Here we explain gradient-based optimization [S2] as an alternative approach that displays better convergence behavior for the present problem. In the following, we assume for simplicity that i) all sites have the same local spin-quantum number $s$, ii) all clusters (total number $Q$) contain the same number of sites $q$, and iii) all non-zero interactions between pairs of spin centers (belonging to the same cluster, or to different clusters) have the same strength in the Heisenberg model, $J_{ij} = 1$. We use indices $i$ and $i'$ for clusters ($i,i' = 1,2,\ldots,Q$), and $p$ and $p'$ for sites in a cluster ($p,p' = 1,2,\ldots,q$).

In Eq. (S2), a Thouless rotation $e^{\hat{Z}}$ relates $|\Phi\rangle$ to an initial guess $|\Phi^0\rangle$,

$$|\Phi\rangle = A e^{\hat{Z}} |\Phi^0\rangle \quad . \tag{S2}$$

$A$ is a normalization constant. This rotation separates into rotations for all individual clusters,

$$|\Phi\rangle = \prod_{i=1}^{Q}\left(A_i e^{\hat{Z}_i}|\Phi_i^0\rangle\right) \quad , \tag{S3}$$

where

$$\hat{Z}_i = \sum_{v \in \text{virt}} \sum_{o \in \text{occ}} Z_{i,vo} \hat{c}^\dagger_{i,v} \hat{c}_{i,o} \ . \tag{S4}$$

The quantum-chemical terminology used in Eq. (S4), occ = "occupied", virt = "virtual", recurs to a fermionic formulation (cf. Ref. [S1]), where $|\Phi^0_i\rangle$ defines a single occupied molecular orbital ($o = 1$) at the respective cluster in terms of a fermionic creation operator,

$$|\Phi_i\rangle \leftrightarrow \hat{c}^\dagger_{i,o} |0_i\rangle, \tag{S5}$$

where $|0_i\rangle$ is the vacuum at cluster $i$. The remaining (orthogonal) $M-1$ states, where $M = (2s+1)^q$ is the dimension of the local Hilbert space, defines a set of virtual orbitals. The $M-1$ optimization parameters $Z_{i,vo}$ are the elements of a complex column vector $\mathbf{Z}_i$ (the Thouless vector). The essence of cPHF consists in minimizing the energy $E$ of the projected state $|\Psi\rangle = \hat{P}|\Phi\rangle$.

As indicated, we find it helpful to draw a conceptual connection to electronic-structure theory by associating the state of a spin cluster with a molecular orbital occupied by a single fermion. This is analogous to our approach to PHF [S1], which can be regarded as a special case of cPHF with cluster size $q = 1$. A fermionic formulation leads to a second-quantized Hamiltonian,

$$\hat{H} = \sum_{lm} t_{lm} \hat{c}^\dagger_l \hat{c}_m + \frac{1}{2} \sum_{klmn} \hat{c}^\dagger_k \hat{c}^\dagger_l \hat{c}_m \hat{c}_n [kn|lm] \ , \tag{S6}$$

parameterized by one- and two-body integrals $t_{lm}$ and $[kn|lm]$, respectively. Projection into the subspace of states that have exactly one fermion per site is implicitly assumed, such that the fermionic Hamiltonian (Eq. (S6)) becomes equivalent to the Heisenberg model, $\hat{H} = \sum_{i<j} J_{ij} \hat{\mathbf{s}}_i \cdot \hat{\mathbf{s}}_j$.

Integrals in Eq. (S6) have a simple block structure. The matrix $\mathbf{t}$ comprising matrix elements $t_{lm}$ is of dimension $(Q \cdot M) \times (Q \cdot M)$, but consists of $Q$ blocks $\mathbf{h}_i$, where $\mathbf{h}_i$ is the Hamiltonian of the $i$-th isolated cluster. We similarly define reduced two-body integrals $[kn|lm]_{pp'}$ (for $p, p' = 1, 2, ..., q$) for the interaction of site $p$ of an arbitrary cluster with site $p'$ of another arbitrary cluster. These integrals are generally zero for most combinations $p, p'$, because only specific site pairs interact. Local spin matrices $\mathbf{s}_{p,\alpha}$ ($p = 1, 2, ..., q$; $\alpha = x, y, z$) are of dimension $M \times M$. The number $Z$ of non-zero entries in $\mathbf{s}_{p,\alpha}$ is independent of $p$ (for $s = \frac{1}{2}$, $Z = M$ for all $\alpha$). For a given $\alpha$, there are thus $Z^2$ combinations of non-zero entries in $\mathbf{s}_{p,\alpha}$ and $\mathbf{s}_{p',\alpha}$, each combination yielding a non-vanishing integral $[kn|lm]_{pp'}$. The value of $[kn|lm]_{pp'}$ is the product of the respective non-zero entries in $\mathbf{s}_{p,\alpha}$ and $\mathbf{s}_{p',\alpha}$ (all couplings are assumed to have the same strength, $J = 1$); $(k, n)$ and $(l, m)$ are the (row, column) indices of the non-zero entries in $\mathbf{s}_{p,\alpha}$ and $\mathbf{s}_{p',\alpha}$, respectively.

The initial guess $|\Phi_i^0\rangle$, that is, an initial set of molecular orbitals (MOs), is provided in terms of expansion coefficients $\mathbf{O}_{i,\text{occ}}^0$ in the uncoupled $|m_1, m_2, ..., m_q\rangle$ basis, where the latter corresponds to an orthonormal atomic-orbital (AO) basis in electronic-structure calculations. $\mathbf{O}_{i,\text{occ}}^0$ is an $M \times 1$ vector, and $\mathbf{O}_i^0 = (\mathbf{O}_{i,\text{occ}}^0, \mathbf{O}_{i,\text{virt}}^0)$ is $M \times M$. We generate $|\Phi_i^0\rangle$ through small random mixing of occupied and virtual cHF orbitals. At each iteration of the optimization process, a pure-state density matrix $\boldsymbol{\rho}_i$ for each cluster can be formed,

$$\boldsymbol{\rho}_i = \mathbf{O}_{i,\text{occ}} \mathbf{O}_{i,\text{occ}}^\dagger . \tag{S7}$$

In the first iteration, we set $\mathbf{Z} = 0$ and $\mathbf{O}_i = \mathbf{O}_i^0$, and calculate (as described below) the energy $E$ of the projected state as well as the global gradient with respect $\mathbf{Z}$.

Eqs. (S8)-(S13) describe the updating of $\mathbf{O}_i$ through a Thouless rotation $\mathbf{Z}_i$ from $\mathbf{O}_i^0$ in each iteration. The number $L_i$ and the lower triangular $(M-1) \times (M-1)$ matrix $\mathbf{M}_i$ are obtained from Eqs. (S8) and (S9) by Cholesky decomposition:

$$1 + \mathbf{Z}_i^T \mathbf{Z}_i^* = L_i L_i^* , \tag{S8}$$

$$1 + \mathbf{Z}_i^* \mathbf{Z}_i^T = \mathbf{M}_i \mathbf{M}_i^\dagger . \tag{S9}$$

Following Eqs. 3.45 and 3.47 in Ref. [S2], we form the intermediate $\breve{\mathbf{O}}$,

$$(\breve{\mathbf{O}}_{i,\text{occ}})_k = (\mathbf{O}_{i,\text{occ}}^0)_k + \sum_{v=1}^{M-1} (\mathbf{Z}_i)_v (\mathbf{O}_{i,\text{virt}}^0)_{kv} , \tag{S10}$$

$$(\breve{\mathbf{O}}_{i,\text{virt}})_{lm} = (\mathbf{O}_{i,\text{virt}}^0)_{lm} - (\mathbf{Z}_i)_m^* (\mathbf{O}_{i,\text{occ}}^0)_l , \tag{S11}$$

and finally obtain the properly orthonormalized (unitary) $\mathbf{O}_i$,

$$\mathbf{O}_{i,\text{occ}} = \frac{1}{L^*} \breve{\mathbf{O}}_{i,\text{occ}} , \tag{S12}$$

S3

$$(\mathbf{O}_{i,\text{virt}})_{lm} = \sum_{k=1}^{M-1} (\mathbf{M}_i^{-1})_{lk} (\breve{\mathbf{O}}_{i,\text{virt}})_{km} \ . \tag{S13}$$

In the following, we explain the calculation of the energy $E$ of the projected state (similar to the PHF algorithm described in Ref. [S1], see also Ref. [S3]) and the calculation of the global gradient vector $\mathbf{G}$ with elements $G_{vo}$ $(o=1)$, defined in Eq. (S14).

$$\begin{aligned}\delta E &= -\sum_{vo} \left( G_{vo} \delta Z_{vo}^* + \text{c.c.} \right) = \\ &-2 \sum_{vo} \left( \text{Re}(G_{vo}) \text{Re}(\delta Z_{vo}) + \text{Im}(G_{vo}) \text{Im}(\delta Z_{vo}) \right) \end{aligned} \tag{S14}$$

As noted above, $\mathbf{O}_i$ mediates a transformation from the AO to the MO basis. Matrices defined in the MO basis carry a tilde, where $\tilde{\boldsymbol{\rho}}_i$ assumes a simple form:

$$\tilde{\boldsymbol{\rho}}_i = \mathbf{O}_i^\dagger \boldsymbol{\rho}_i \mathbf{O}_i = \begin{pmatrix} \mathbf{1} & 0 \\ 0 & 0 \end{pmatrix} \ . \tag{S15}$$

In Eq. (S15), the unit matrix $\mathbf{1}$ is $1 \times 1$, because there is only one occupied MO per site. For spin projection, the grid weights $t(\Omega)$ [an Euler-angle triplet $\Omega = (\alpha, \beta, \gamma)$ defines a grid point], are combined with Wigner $D$-matrix elements for all combinations of magnetic quantum numbers $m$ and $k$ [S3],

$$x_{mk}(\Omega) = t(\Omega) D_{mk}^{S*}(\Omega) \ . \tag{S16}$$

For combined S- and PG-projection, $\Lambda$ denotes a combination $\hat{R}_\Lambda = \hat{R}_\Omega \hat{R}_g$ of a spin-rotation $\hat{R}_\Omega$ and a site permutation $\hat{R}_g$. The loop over grid points $\Lambda = (\Omega, g)$ thus comprises two nested loops, for $\Omega$ and $g$. In the spin-rotation matrix $\mathbf{R}_\Omega$ (the matrix representation of $\hat{R}_\Omega$),

$$\mathbf{R}_\Omega = \exp(-i\alpha \boldsymbol{\tau}_z) \times \exp(-i\beta \boldsymbol{\tau}_y) \times \exp(-i\gamma \boldsymbol{\tau}_z) , \tag{S17}$$

$\boldsymbol{\tau} = (\boldsymbol{\tau}_x, \boldsymbol{\tau}_y, \boldsymbol{\tau}_z)$ is the total-spin vector of an isolated cluster, $\boldsymbol{\tau}_\alpha = \sum_{p=1}^{q} \mathbf{s}_{p,\alpha}$.

The PG-operation $\hat{R}_g$ converts cluster $i$ into cluster $g(i)$, which may be associated with a permutation among sites within the clusters (internal spin permutations). Take a dimerized ($q=2$) symmetric $N=6$ spin ring as an example (Figure S1). The cyclic operation $\hat{C}_3$ carries center $p=1$ of cluster $i=1$ into $p'=1$ of $g(i)=2$, etc., and thus does not cause internal permutations, see Figure S1a.

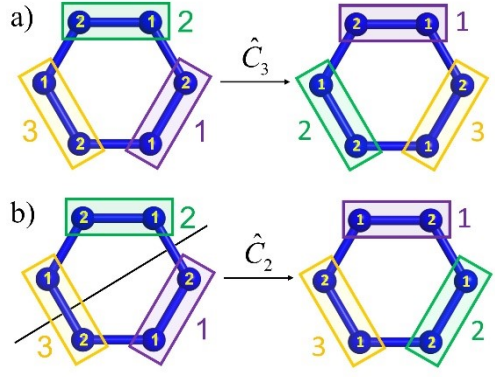

Figure S1: The cyclic point-group operation $\hat{C}_3$ keeps the numbering of spins in the clusters intact (a), but a vertical rotation causes internal permutations (b).

On the other hand, the vertical two-fold rotation $\hat{C}_2$ illustrated in Figure S1b exchanges site pairs in all three clusters. In other words, the $\hat{C}_2$ operation is associated with internal permutations.

With $\hat{R}_g$ causing an internal spin permutation $\bar{g}(i)$ in cluster $g(i)$, the single-cluster block $\tilde{\mathbf{R}}_{i,\Lambda}$ of the combined spin-rotation/PG-operation becomes:

$$\tilde{\mathbf{R}}_{i,\Lambda} = \mathbf{O}_i^\dagger \mathbf{R}_\Omega \mathbf{P}_{\bar{g}(i)} \mathbf{O}_{g(i)}. \tag{S18}$$

$\mathbf{P}_{\bar{g}(i)}$ is the internal permutation operator. As an example, for $q=2$, $\bar{g}(i)$ can have only two values (exchange, or no exchange). For $s=\tfrac{1}{2}$, the exchange operator has the simple representation $\mathbf{P} = \tfrac{1}{2}(\mathbf{1} + 4\mathbf{s}_1 \cdot \mathbf{s}_2)$ [S4]. Permutation operators for $s > \tfrac{1}{2}$ were derived in Ref. [S5].

$\tilde{\mathbf{R}}_{i,\Lambda}$ has four blocks,

$$\tilde{\mathbf{R}}_{i,\Lambda} = \begin{pmatrix} \tilde{\mathbf{R}}_{i,\Lambda}^{oo} & \tilde{\mathbf{R}}_{i,\Lambda}^{ov} \\ \tilde{\mathbf{R}}_{i,\Lambda}^{vo} & \tilde{\mathbf{R}}_{i,\Lambda}^{vv} \end{pmatrix}, \tag{S19}$$

which are superscripted by o ("occupied") or v ("virtual"). Only the first column of $\tilde{\mathbf{R}}_{i,\Lambda}$, consisting of $\tilde{\mathbf{R}}_{i,\Lambda}^{oo}$ ($1 \times 1$) and $\tilde{\mathbf{R}}_{i,\Lambda}^{vo}$ [$(M-1) \times 1$], is needed to form i) the rotated overlap, $Q_\Lambda \equiv \langle \Phi | \hat{R}_\Omega \hat{R}_g | \Phi \rangle$, from a product of the rotated overlaps of all clusters, Eq. (S20),

$$Q_\Lambda = \prod_i \tilde{\mathbf{R}}_{i,\Lambda}^{oo}, \tag{S20}$$

and ii) the rotated transition-density matrices $\tilde{\boldsymbol{\rho}}_{i,\Lambda}$,

S5

$$\tilde{\boldsymbol{\rho}}_{i,\Lambda} = \begin{pmatrix} 1 & 0 \\ \tilde{\mathbf{R}}_{i,\Lambda}^{vo} \left[ \tilde{\mathbf{R}}_{i,\Lambda}^{oo} \right]^{-1} & 0 \end{pmatrix}. \tag{S21}$$

The latter are transformed back to the AO basis,

$$\boldsymbol{\rho}_{i,\Lambda} = \mathbf{O}_i \tilde{\boldsymbol{\rho}}_{i,\Lambda} \mathbf{O}_i^{\dagger}, \tag{S22}$$

and contracted with four-index interaction integrals (defined in the AO basis), Eq. (S23),

$$(\Delta \mathbf{G}_{i,\Lambda})_{kl} = \sum_{mn} \left[ kl \mid mn \right]_{pp'} (\boldsymbol{\rho}_{i',\Lambda})_{mn}. \tag{S23}$$

The perturbation tensor $\mathbf{G}_{i,\Lambda}$ is the sum of all $\Delta \mathbf{G}_{i,\Lambda}$ increments, as summarized in the following scheme.

```
for i = 1 : Q (sum over clusters)
    for p = 1 : q (sum over sites in cluster i)
        for all (i', p') interacting with (i, p) (external bonds, i.e., i ≠ i')
            (ΔG_{i,Λ})_{kl} = Σ_{mn} [kl|mn]_{pp'} (ρ_{i',Λ})_{mn}
            (G_{i,Λ})_{kl} = (G_{i,Λ})_{kl} + (ΔG_{i,Λ})_{kl}
        end
    end
end
```

We define $\mathbf{F}_{i,\Lambda}$ as the sum of the local cluster Hamiltonian and the perturbation tensor,

$$\mathbf{F}_{i,\Lambda} = \mathbf{h}_i + \mathbf{G}_{i,\Lambda}. \tag{S24}$$

The quantity $V_\Lambda$ is the trace of the sum $\mathbf{h}_i + \mathbf{F}_{i,\Lambda}\boldsymbol{\rho}_{i,\Lambda}$, taken over all blocks:

$$V_\Lambda = \sum_i \mathrm{Tr}\left(\mathbf{h}_i + \mathbf{F}_{i,\Lambda}\boldsymbol{\rho}_{i,\Lambda}\right). \tag{S25}$$

Back in the MO basis, $\tilde{\mathbf{F}}_{i,\Lambda} = \mathbf{O}_i^{\dagger} \mathbf{F}_{i,\Lambda} \mathbf{O}_i$. At each grid point $\Lambda$ the following quantities are incremented (for Eqs. (S28) and (S29), see Eqs. 3.37 and 3.38 in Ref. [S2]):

$$W_{mk} \mathrel{+}= w_{\Lambda,mk}, \tag{S26}$$

$$H_{mk} \mathrel{+}= \tfrac{1}{2} w_{\Lambda,mk} V_\Lambda, \tag{S27}$$

$$\tilde{\mathbb{R}}_{i,mk} \mathrel{+}= w_{\Lambda,mk} \tilde{\mathbf{r}}_{i,\Lambda}, \tag{S28}$$

$$\tilde{\mathbb{T}}_{i,mk} \mathrel{+}= w_{\Lambda,mk} \left[ \tfrac{1}{2} V_\Lambda \tilde{\mathbf{r}}_{i,\Lambda} + (\mathbf{1} - \tilde{\boldsymbol{\rho}}_{i,\Lambda}) \tilde{\mathbf{F}}_{i,\Lambda} \tilde{\mathbf{r}}_{i,\Lambda} \right], \tag{S29}$$

where $\tilde{\mathbf{r}}_{i,\Lambda}$ denotes the first column of $\tilde{\boldsymbol{\rho}}_{i,\Lambda}$ and $w_{\Lambda,mk}$ is defined in Eq. (S30),

$$w_{\Lambda,mk} = x_{mk}(\Omega)\chi_\Gamma^*(g)Q_\Lambda \ , \tag{S30}$$

and $\chi_\Gamma^*(g)$ is an element of the character vector (the PG-projector is $\hat{P}_\Gamma = \tfrac{1}{h}\sum_{g=1}^{h}\chi_\Gamma^*(g)\hat{R}_g$ ).

The generalized eigenvalue-problem is solved for the lowest energy $E$,

$$\mathbf{Hf} = E\mathbf{Wf} \ , \tag{S31}$$

under the normalization constraint $\mathbf{f}^\dagger\mathbf{Wf}=1$. We assemble the local gradient $\tilde{\mathbf{g}}_i$ from the $\tilde{\mathbb{R}}_{i,mk}$ and $\tilde{\mathbb{T}}_{i,mk}$ sets of vectors,

$$\tilde{\mathbf{g}}_i = -\sum_{MK} f_m^* f_k \left( \left[\tilde{\mathbb{T}}_{i,mk}\right]_{\text{virt}} - E\left[\tilde{\mathbb{R}}_{i,mk}\right]_{\text{virt}} \right) \ , \tag{S32}$$

where the notation $\left[\tilde{\mathbb{T}}_{i,mk}\right]_{\text{virt}}$ signifies that the first element of the vector $\tilde{\mathbb{T}}_{i,mk}$ is excluded, that is, $\tilde{\mathbf{g}}_i$ is $(M-1)\times 1$. The global gradient $\mathbf{G}_i$ is obtained by transforming to the MO basis of the initial guess [S2],

$$\mathbf{G}_i = \frac{1}{L_i^*}(\mathbf{M}_i^T)^{-1}\tilde{\mathbf{g}}_i \ . \tag{S33}$$

The $\mathbf{G}_i$ for all $Q$ clusters are concatenated into $\mathbf{G}$, which is $Q\cdot(M-1)\times 1$. The energy $E$ of the projected state, the initial guess $\mathbf{O}^0$ and the global gradient $\mathbf{G}$ are passed to the `fminunc` function in `Matlab`. The separation into real and imaginary parts (Eq. (S14)) is required for `fminunc`, which optimizes with respect to a set of real variables.

Note that for cPHF with a UHF (instead of a GHF reference) we first run a few iterations with an SCF algorithm to determine the local spin-projections on the $z$-axis, $m_i$. With the initial guess thus established, we switch to gradient-based optimization, setting those elements of the gradient to zero which would change $m_i$. In other words, in gradient-based optimization, all $m_i$ numbers are frozen at their values in the initial guess $\mathbf{O}^0$.

## 2. Spin-pair correlation functions (SPCFs)

The calculation of SPCFs is analogous to the evaluation of the energy. A double-integration over the spin-projection grid can be avoided, because $\hat{\mathbf{s}}_i \cdot \hat{\mathbf{s}}_j$ is a spin scalar which commutes with the (Hermitian and idempotent) spin-projection operator. For PG-projection, we consider only one-dimensional representations $\Gamma$. Then only the totally symmetric part $(\hat{\mathbf{s}}_i \cdot \hat{\mathbf{s}}_j)_{\Gamma_1}$, Eq. (S34),

$$(\hat{\mathbf{s}}_i \cdot \hat{\mathbf{s}}_j)_{\Gamma_1} = \frac{1}{h}\sum_{g=1}^{h} \hat{R}_g^{\dagger}(\hat{\mathbf{s}}_i \cdot \hat{\mathbf{s}}_j)\hat{R}_g , \qquad (S34)$$

contributes to $\langle\Phi|\hat{P}_\Gamma^{\dagger}(\hat{\mathbf{s}}_i \cdot \hat{\mathbf{s}}_j)\hat{P}_\Gamma|\Phi\rangle$. Overall, a single summation/integration is required to evaluate SPCFs for PGSGHF wave functions [S1],

$$\langle\Phi|\hat{P}_S^{\dagger}\hat{P}_\Gamma^{\dagger}(\hat{\mathbf{s}}_i \cdot \hat{\mathbf{s}}_j)\hat{P}_S\hat{P}_\Gamma|\Phi\rangle = \langle\Phi|(\hat{\mathbf{s}}_i \cdot \hat{\mathbf{s}}_j)_{\Gamma_1}\hat{P}_S\hat{P}_\Gamma|\Phi\rangle . \qquad (S35)$$

The evaluation of the expectation value for each term $\hat{\mathbf{s}}_l \cdot \hat{\mathbf{s}}_m$ occurring in the symmetrized operator $(\hat{\mathbf{s}}_i \cdot \hat{\mathbf{s}}_j)_{\Gamma_1}$ depends on whether sites $l$ and $m$ are in the same cluster (corresponding to a single-particle term) or in different clusters (two-particle term).

### 3. cPHF energies for spin rings

Table 1: Singlet and triplet energies and the gap $\Delta E_{ST}$ in $s=\frac{1}{2}$ rings with $N$ sites and cluster size $q$. SUHF predictions are compared to exact results (from Table V.1 in Ref. [S6]).

| $q$ | | $N$ | | | | |
|---|---|---|---|---|---|---|
| | | 6 | 12 | 18 | 24 | 30 |
| 2 | $E_S$ | -2.6514 | -4.8770 | -7.1335 | -9.3914 | -11.6485 |
| | $E_T$ | -1.8956 | -4.3114 | -6.6414 | -8.9434 | -11.2237 |
| | $\Delta E_{ST}$ | 0.756 | 0.566 | 0.492 | 0.448 | 0.425 |
| 6 | $E_S$ | -2.8028 | -5.3482 | -7.7332 | -10.2071 | -12.6941 |
| | $E_T$ | -2.1180 | -4.7874 | -7.3492 | -9.8772 | -12.3954 |
| | $\Delta E_{ST}$ | 0.685 | 0.561 | 0.384 | 0.330 | 0.299 |
| Exact | $E_S$ | -2.803 | -5.387 | -8.023 | -10.670 | -13.322 |
| Exact | $\Delta E_{ST}$ | 0.685 | 0.356 | 0.241 | 0.183 | 0.147 |

Table 2: Singlet and triplet energies and the gap $\Delta E_{ST}$ in $s = \frac{1}{2}$ rings with $N$ sites and cluster size $q$. SGHF predictions are compared to exact results (from Table V.1 in Ref. [S6]).

| $q$ | | $N$ | | | | |
|---|---|---|---|---|---|---|
| | | 6 | 12 | 18 | 24 | 30 |
| 2 | $E_S$ | -2.8028 | -5.0625 | -7.3603 | -9.6589 | -11.9416 |
| | $E_T$ | -2.1180 | -4.5485 | -6.8696 | -9.1446 | -11.4453 |
| | $\Delta E_{ST}$ | 0.685 | 0.514 | 0.491 | 0.514 | 0.496 |
| 6 | $E_S$ | -2.8028 | -5.3768 | -7.9641 | -10.4728 | -12.9231 |
| | $E_T$ | -2.1180 | -5.0090 | -7.6042 | -10.1411 | -12.6090 |
| | $\Delta E_{ST}$ | 0.685 | 0.368 | 0.360 | 0.332 | 0.314 |
| Exact | $E_S$ | -2.803 | -5.387 | -8.023 | -10.670 | -13.322 |
| Exact | $\Delta E_{ST}$ | 0.685 | 0.356 | 0.241 | 0.183 | 0.147 |

Table 3: Singlet and triplet energies and the gap $\Delta E_{ST}$ in $s = \frac{1}{2}$ rings with $N$ sites and cluster size $q$. $D_Q$SGHF predictions ($Q = N/q$) are compared to exact results (from Table V.1 in Ref. [S6]).

| $q$ | | $N$ | | | | |
|---|---|---|---|---|---|---|
| | | 6 | 12 | 18 | 24 | 30 |
| 2 | $E_S$ | -2.8028 | -5.3710 | -7.8905 | -10.3945 | -12.8677 |
| | $E_T$ | -2.1180 | -5.0104 | -7.5544 | -10.0287 | -12.4834 |
| | $\Delta E_{ST}$ | 0.685 | 0.361 | 0.336 | 0.366 | 0.384 |
| 6 | $E_S$ | -2.8028 | -5.3874 | -8.0224 | -10.6501 | -13.2762 |
| | $E_T$ | -2.1180 | -5.0315 | -7.7782 | -10.4356 | -13.0870 |
| | $\Delta E_{ST}$ | 0.685 | 0.356 | 0.244 | 0.215 | 0.189 |
| Exact | $E_S$ | -2.803 | -5.387 | -8.023 | -10.670 | -13.322 |
| Exact | $\Delta E_{ST}$ | 0.685 | 0.356 | 0.241 | 0.183 | 0.147 |